\documentclass[prc,showpacs,floatfix,]{revtex4}
\usepackage{graphicx,subfigure}
\usepackage{comment}
\usepackage{amsmath, amsthm, amssymb}
\usepackage{bm}

\def\nuc#1#2{\relax\ifmmode{}^{#1}{\protect{#2}}\else${}^{#1}$#2\fi}

\newcommand {\la} {\langle}
\newcommand {\ra} {\rangle}
\newcommand {\beq} {\begin{eqnarray}}
\newcommand {\eeq} {\end{eqnarray}}
\newcommand {\eeqn} [1] {\label{#1} \end{eqnarray}}%
\newcommand {\eol} {\nonumber \\}
\newcommand {\ve} [1] {\mbox{\boldmath $#1$}}

\begin{document}
\graphicspath{{figures/}}

\title{Antisymmetrized, translationally invariant theory of the nucleon optical potential.}
 \author{R.C. Johnson}
 \affiliation{
  Department of Physics,
 Faculty of Engineering and Physical Sciences,
 University of Surrey,
 Guildford, Surrey GU2 7XH, United Kingdom}

\date{\today}
\begin{abstract}
Earlier work showed how a nucleon optical model wave function could be defined as a projection of a many-nucleon scattering state within a translationally invariant second quantised many-body theory. In this paper an optical potential operator that generates this optical model wave function is defined through a particular off-shell extension of the elastic transition operator. The theory is express explicitly in terms of the many-nucleon Hamiltonian in a mixed representation in which localised target nucleus states feature. No reference to a mean-field concept is involved in the definition.  It is shown that the resulting optical model operator satisfies the requirements of rotational invariance and translational invariance and has standard behaviour under the time reversal transformation. 

The contributions to the optical potential from two different  exchange mechanisms are expressed in terms of an effective Hamiltonian involving a nucleon-number conserving one-body interaction. In the weak-binding limit the method reduces to a a version of Feshbach's projection operator formulation of the optical potential with a truncated nucleon-nucleon potential including exchange terms and recoil corrections.

Definitions of the nucleon single-particle Green's function and the corresponding  Dyson self-energy  modified by corrections for translational invariance are presented and different definitions of the optical potential operator are compared. 
\end{abstract}
\maketitle
\section{Introduction.}\label{intro}
Theories describing nucleon scattering from an $A$-nucleon target in terms of  fundamental 2- and 3-body inter-nucleon interactions are of considerable current interest. An extensive recent review of the optical potential approach is given in \cite{DickhoffCharity2018}. Apart from the case of very light targets all theories cited in \cite{DickhoffCharity2018} ignore the full implications of translational invariance. 

  As a step towards correcting this situation, earlier work \cite{Johnson17} showed how a nucleon optical model wave function could be defined as a projection of a many-nucleon scattering state within a translationally invariant second quantised many-body theory. In sub-Sections \ref{optwf}and \ref{source} the results of \cite{Johnson17} are reviewed briefly. In Sections \ref{optdef}-\ref{nonuniqueVopt}, in a  natural development  of \cite{Johnson17}, an optical potential operator that generates the optical model wave function is defined in terms of a particular off-shell extension of the elastic transition operator and expressed explicitly in terms of the many-nucleon Hamiltonian. No reference to a mean-field concept is involved in this definition which is shown to produce an optical potential that satisfies the requirements of rotational invariance and translational invariance and has a standard behaviour under the time reversal transformation.

In sub-Sections \ref{physinterp}-\ref{summ1} the new definition is used to distinguish the contributions from two different exchange mechanisms to the optical potential, knock-out exchange and heavy particle stripping. For an $A$-nucleon target these two contributions are expressed in terms of an effective Hamiltonian involving a nucleon number conserving interaction acting within $A$-nucleon and $(A-1)$-nucleon sub-spaces.  Section \ref{HPS term} shows how the heavy particle stripping term is related to the hole-term in the Lehmann, Symanzik and Zimmermann representation of the transition operator \cite{Villars1967}. In Section \ref{quasi free} it is shown that in the weak-binding limit the theory reduces to a modified version of Feshbach's original theory of the optical potential \cite{Fes58}.

In Section \ref{GreenfunctionDyson} a modified single-particle Green's function and Dyson self-energy are derived. The associated definition of an optical potential is compared with that given in Section \ref{opt model} . 

Concluding discussions can be found in Section \ref{DisCons} and acknowledgements in Section \ref{acks}. Details of some of the derivations are collected in an Appendices A-E.

The inclusion of recoil effects in many-body theories of the optical potential was discussed by Redish and Villars \cite{RedishVillars1970}. Their work  derived corrections to systematic perturbation methods about a mean field. Although much of the analysis in this paper shares with \cite{RedishVillars1970} the use of  techniques developed in \cite{Villars1967}, no attempt is made here to develop a perturbation theory. The motivation here is rather to write down a theory of the optical potential that explicitly satisfies anti-symmetry and translational invariance requirements and brings out the physical content in a way that bridges the gap between many-body theory and standard nuclear reaction theory ideas.

\section{Theory of the nucleon optical potential.}\label{opt model}
 \subsection{The optical model wavefunction.}\label{optwf}
In \cite{Johnson17}, the optical model wave function, $ \xi_{E,\ve{k}_0}^\epsilon(\ve{r})$, corresponding to the elastic scattering of a nucleon of momentum $\ve{k}_0$ in the overall c.m. system by an $A$-nucleon target in its ground state was formally defined as a matrix element between many-nucleon states  in Fock-space through the formula
 \beq \xi_{E,\ve{k}_0}^\epsilon(\ve{r})=\la\la \Psi(0, \ve{x}=0) \mid \psi(\ve{r})\mid \Psi_{E,\ve{k}_0}^\epsilon\ra\ra. \label{xiEepsdef}\eeq
The operator $\psi(\ve{r})$ destroys a nucleon at a point labelled $\ve{r}$. The notation $\ve{r}$ will be taken to include  spin and iso-spin coordinates of a nucleon unless it is obvious that only space coordinates are referred to by the context. The notation $\mid\,\ra\ra$ and $\la\la\,\mid$  denotes kets and bras in Fock-space.
 
 The ground state energy will be taken to be the zero of energy so that total c.m. energy $E$ is related to $\ve{k}_0$ on the energy shell by
\beq E=\frac{\hbar^2k^2_0}{2\mu_{mA}},\label{E1}\eeq
where $\mu_{mA}$ is the nucleon-target reduce mass
\beq \mu_{mA}=\frac{A}{(A+1)}m.\label{mumA}\eeq
In order to make subsequent formulae have a simpler appearance the difference between neutron and proton rest masses will be ignored.

 The ket on the right of eq.(\ref{xiEepsdef})  is the Fock-space scattering state\cite{Johnson17}
\beq\mid \Psi_{E,\ve{k}_0}^\epsilon\ra\ra =\frac{\imath \epsilon}{E-H+\imath \epsilon}(2\pi)^{3/2}a^\dagger_{\ve{k}_0}\mid -\ve{k}_0,\psi_0 \ra\ra. \label{Psi1scatt2}\eeq 
where $H$ is the many-nucleon Hamiltonian operator in Fock-space.

In the limit $\epsilon \rightarrow 0^+$ the ket $\mid \Psi_{E,\ve{k}_0}^\epsilon\ra\ra$ describes an antisymmetric  $(A+1)$-nucleon scattering state of total momentum zero in the overall c.m. system. The incident channel  has an incident nucleon with momentum $\ve{k}_0$. In this channel the $A$-nucleon  target has a total momentum $-\ve{k}_0$ and is in its ground state $\psi_0$. All other channel components of  $\mid \Psi_{E,\ve{k}_0}^\epsilon\ra\ra$ have purely outgoing waves asymptotically.  It will be assumed that all Coulomb interactions are screened at large separations. The factor $(2\pi)^{3/2}$ arises because $a^\dagger_{\ve{k}_0}$ creates a normalised plane wave state, whereas scattering states are conventionally normalised to an incoming plane wave of unit amplitude.

The bra $\la\la \Psi(0, \ve{x}=0) \mid$ on the right of eq.(\ref{xiEepsdef}) describes a state in which the target is in its ground state and its c.m. is located at the origin of coordinates, $\ve{x}=0$. This  is one of the complete set of A-nucleon Fock-space states, $\mid \Psi(n, \ve{x}) \rangle \rangle$,  formed from  an intrinsic state, $\psi_n(\ve{r}_1, \dots, \ve{r}_A)$,  and  and having  a c.m. located at position $\ve{x}$. Explicitly \cite{Johnson17}
\beq 
 \mid \Psi(n, \ve{x}) \rangle \rangle =\frac{1}{\sqrt{A!}}\int d\ve{r}_1\, d\ve{r}_2 \dots d\ve{r}_A \delta(\ve{R}_A-\ve{x})\psi_n(\ve{r}_1, \dots, \ve{r}_A)\psi^\dagger(\ve{r}_A)\dots \psi^\dagger(\ve{r}_1)\mid 0 \rangle \rangle,\eol&&\label{Psi0xF}\eeq
 where
 \beq \ve{R}_A=\frac{(\ve{r}_1+ \dots+\ve{r}_A)}{A}.\label{RA}\eeq 
 The state $\psi_n$ is an eigenstate of the intrinsic part of the $A$-nucleon Hamiltonian $H_A$
 \beq H_A-\frac{(\ve{P})^2}{2Am},  \label{Hint}\eeq
 where $\ve{P}$ is the momentum operator in Fock-space
 \beq\ve{P} =\hbar \int d\ve{k}\, \ve{k}\,a^\dagger_{\ve{k}}\,a^\dagger_{\ve{k}}.\label{Pdef}\eeq
 The state $\psi_n(\ve{r}_1, \dots, \ve{r}_A)$ is an eigenfunction of $\ve{P}$ with eigenvalue zero. The ket $\mid -\ve{k}_0,\psi_0 \ra\ra$ in eq.(\ref{Psi1scatt2}) is an eigenfunction of $\ve{P}$ with eigenvalue $-\ve{k}_0$.
 
 In eq.(\ref{xiEepsdef}) the argument $\ve{r}$ of $ \xi_{E,\ve{k}_0}^\epsilon(\ve{r})$ can be interpreted as the position of the incident nucleon relative to the c.m. of the target, although of course in the scattering state the target c.m. is not at rest in the overall c.m. system. This is reflected in the complete uncertainty of the momentum of the bra $\la\la \Psi(0, \ve{x}=0) \mid \psi(\ve{r})$. The translational invariance of $H$ means that only the zero total momentum component of this bra contributes to the matrix element (\ref{xiEepsdef}).
 
 \subsection{Source equation for the optical model wave function.}\label{source}
 The optical model wave function defined by eq.(\ref{xiEepsdef}) satisfies \cite{Johnson17}
 \beq(\frac{\hbar^2k^2_0}{2\mu_{mA}}+\imath \epsilon+\frac{\hbar^2}{2\mu_{mA}}\nabla^2_{\ve{r}}) \xi_{E,\ve{k}_0}^\epsilon(\ve{r})&=&F^\epsilon_{\ve{k}_0}(\ve{r})+\imath \epsilon \exp(\imath  \ve{k}_0.\ve{r}), \label{source2}\eeq
where
 \beq F^\epsilon_{\ve{k}_0}(\ve{r})= \la\la \Psi(0,\ve{x}=0) \mid[ \psi(\ve{r}),V]_-\mid \Psi_{E,\ve{k}_0}^\epsilon\ra\ra.\label{Fk0r}\eeq
The form eq.(\ref{source2}) takes in the $\epsilon \rightarrow 0$ is derived in Section V of \cite{Johnson17}. A straightforward modification of the argument given there gives the finite $\epsilon $ version, eq. (\ref{source2}).

The  optical potential operator $\hat{V}^\mathrm{opt}(E)$ will be defined as an operator in barycentric space ($B$-space), the space of a fictitious particle of mass $\mu_{mA}$, spin-$1/2$, position operator $\hat{\ve{r}}$ and momentum operator $\hat{\ve{p}}=-\imath \hbar \nabla_r$, that enables eq.(\ref{source2}) to be written in the equivalent form
 \beq(\frac{\hbar^2k^2_0}{2\mu_{mA}}+\imath \epsilon+\frac{\hbar^2}{2\mu_{mA}}\nabla^2_{\ve{r}}) \xi_{E,\ve{k}_0}^\epsilon(\ve{r})&=&\hat{V}^\mathrm{opt}(E) \xi_{E,\ve{k}_0}^\epsilon(\ve{r})+\imath \epsilon \exp(\imath  \ve{k}_0.\ve{r}). \eol &&\label{source3}\eeq
 The "hat" notation over a quantity will be use to denote operators in $B$-space, as opposed to the bare-headed operators of Fock-space.
 
 In general $V^\mathrm{opt}(E)$ will be a non-local operator  with matrix elements in the $\ve{r}$-representation in $B$-space  $\hat{V}^\mathrm{opt}(E;\ve{r},\ve{r}')$, so that the first term on the right in eq.(\ref{source2}) has the form
 \beq \hat{V}^\mathrm{opt}(E) \xi_{E,\ve{k}_0}^\epsilon(\ve{r})=\int d\ve{r}'\hat{V}^\mathrm{opt}(E;\ve{r},\ve{r}') \xi_{E,\ve{k}_0}^\epsilon(\ve{r}'). \eol &&\label{source4}\eeq
 To achieve the identification of this operator, eq.(\ref{source2}) must be written as a relation between operators and kets in $B$-space. This can be achieved in many ways. The choice made here results in an operator that conserves the total angular momentum  in $B$-space and has features which permits a transparent representation of the physical processes involved. 

 $F^\epsilon_{\ve{k}_0}(\ve{r})$, given in eq.(\ref{Fk0r}), is first rewritten using an expression for the scattering state, eq.(\ref{Psi1scatt2}), as a plane wave in the incident channel plus  outgoing wave components. This step uses the Green's function identity \cite{Villars1967}, eq.(3.42), page 315,
  \beq  G(z)a^{\dagger}_{\ve{k}_0}&=& a^{\dagger}_{\ve{k}_0}G(z-\epsilon_{k_0})+G(z)[V,a^{\dagger}_{\ve{k}_0}]_-
G(z-\epsilon_{k_0})\label{Villars1}\eeq
 where $G(z)=\frac{1}{(z -H)}$ for arbitrary complex $z$ and 
 \beq \epsilon_{\ve{k}_0}=\frac{\hbar k_0^2}{2m}.\label{epsk0}\eeq
 It has been assumed that $H$ can be expressed as the sum of a nucleon kinetic energy term $T$ and an inter-nucleon potential energy term $V$.
 
 Acting on the ket $\mid-\ve{k}_0, \psi_0\ra\ra$ 
 \beq \imath \epsilon G(E+\imath \epsilon-\epsilon_{k_0})\mid-\ve{k}_0, \psi_0\ra\ra=\imath \epsilon\frac{1}{(E+\imath \epsilon-\epsilon_{k_0}-\frac{1}{A}\epsilon_{k_0})}\mid-\ve{k}_0, \psi_0\ra\ra=\mid-\ve{k}_0, \psi_0\ra\ra,\label{Gk0psi0}\eeq
 when eq.(\ref{E1}) is satisfied. The definition (\ref{Psi1scatt2}) can therefore be rewritten
 \beq (2\pi)^{-3/2}\mid \Psi_{E,\ve{k}_0}^\epsilon\ra\ra=a^{\dagger}_{\ve{k}_0}\mid  -\ve{k}_0,  \psi_{0} \ra\ra+G(E+\imath \epsilon)[V,a^{\dagger}_{\ve{k}_0}]_- \mid-\ve{k}_0, \psi_0\ra\ra, \label{Psidev}\eeq
 and the expression (\ref{Fk0r}) for $F^\epsilon_{\ve{k}_0}(\ve{r})$ becomes 
 \beq F^\epsilon_{\ve{k}_0}(\ve{r})&&= (2\pi)^{3/2}[\la\la \Psi(0,\ve{x}=0) \mid[ \psi(\ve{r}),V]_-a^{\dagger}_{\ve{k}_0}\mid  -\ve{k}_0,  \psi_{0} \ra\ra\eol&&+\la\la \Psi(0,\ve{x}=0) \mid[ \psi(\ve{r}),V]_-G(E+\imath \epsilon)[V,a^{\dagger}_{\ve{k}_0}]_- \mid-\ve{k}_0, \psi_0\ra\ra].\label{Fk0r2}\eeq
The first term on the right involves groundstate-groundstate matrix elements of $V$. In the second term the groundstate is coupled to a complete set of intermediate $(A+1)$-nucleon  states.
       
       Recall that by definition in the state $\mid -\ve{k}_0,\psi_0 \ra\ra$ the $A$-nucleon c.m. is in a plane-wave state of momentum $-\ve{k}_0$ and unit amplitude and hence
       \beq \mid -\ve{k}_0,\psi_0 \ra\ra=\int d\ve{x}\exp(-\imath  \ve{k}_0.\ve{x})\mid \Psi(0,\ve{x}) \ra\ra, \label{k0x}\eeq
 where $\mid \Psi(0,\ve{x}) \ra\ra$ is one of the states defined in eq.(\ref{Psi0xF}).  Eq.(\ref{Fk0r2})  can now be developed as 
\beq F^\epsilon_{\ve{k}_0}(\ve{r})&&= (2\pi)^{3/2}\int d\ve{x}\exp(-\imath  \ve{k}_0.\ve{x})[\la\la \Psi(0,\ve{x}=0) \mid[ \psi(\ve{r}),V]_-a^{\dagger}_{\ve{k}_0}\mid \Psi(0,\ve{x}) \ra\ra\eol&&+\la\la \Psi(0,\ve{x}=0) \mid[ \psi(\ve{r}),V]_-G(E+\imath \epsilon)[V,a^{\dagger}_{\ve{k}_0}]_- \mid \Psi(0,\ve{x}) \ra\ra]\eol&&=\int d\ve{r}'\hat{\mathcal{T}}_{0,0}(E+\imath \epsilon; \ve{r},\ve{r}')\exp(\imath  \ve{k}_0.\ve{r}') \eol &&\label{Fk0r3}\eeq
where 
\beq\hat{\mathcal{T}}_{0,0}(E+\imath \epsilon; \ve{r},\ve{r}')&&=\int d\ve{x}[\la\la \Psi(0,\ve{x}=0) \mid[ \psi(\ve{r}),V]_-\psi^{\dagger}(\ve{r}'+\ve{x})\mid \Psi(0,\ve{x}) \ra\ra\eol&&+\la\la \Psi(0,\ve{x}=0) \mid[ \psi(\ve{r}),V]_-G(E+\imath \epsilon)[V,\psi^{\dagger}(\ve{r}'+\ve{x})]_- \mid \Psi(0,\ve{x}) \ra\ra]. \label{TE00}\eeq
The role of the integration over $\ve{x}$ is to pick out the momentum zero component of the ket $\psi^\dagger(\ve{r}'+\ve{x})  \mid \Psi(0,\ve{x}) \ra\ra$. This is particularly transparent when the properties of the total momentum operator $\ve{P}$ are used to write
\beq\int d\ve{x} \,\psi^\dagger(\ve{r}'+\ve{x})  \mid \Psi(0,\ve{x}) \ra\ra&&=\int d\ve{x}\,\exp(-\imath \ve{x}.\ve{P})\psi^\dagger(\ve{r}')  \mid \Psi(0,\ve{x}=0) \ra\ra \eol &&=(2\pi)^3\delta(\ve{P})\psi^\dagger(\ve{r}')  \mid \Psi(0,\ve{x}=0) \ra\ra.\label{PpsiPsi}\eeq
  Eq.(\ref{TE00}) defines a fully-off shell transition matrix that is independent of the direction of the incident momentum $\ve{k}_0$. This feature will be essential for the angular-momentum conserving property of the optical model defined in the next section.
   
   All the terms on the right in eq.(\ref{TE00}) involve matrix elements between spatially localised $A$-nucleon states. The range of integration of the c.m. coordinate $\ve{x}$ is also strongly limited in realistic physical situations. The range of non-locality associated with the distance $\mid \ve{r}-\ve{r}'\mid $ is limited by the range of non-locality of $V$ and the spatial dimensions of the target. The range of the variable $\ve{x}$, \emph{i.e.}, the change in the c.m. position of the target during the collision,  can be estimated from   
  \beq (\mathrm{Speed\, of\, c.m.})\times&& \!\!\!\!\!\!(\mathrm{Time\, for\, nucleon}\,\mathrm{to\,travel}\, \mid \ve{r}-\ve{r}'\mid)\eol 
  &&=\frac{\hbar k_0}{Am}\times \frac{\mid \ve{r}-\ve{r}'\mid}{\hbar k_0/m} \eol
  &&= \frac{\mid \ve{r}-\ve{r}'\mid}{A}.\label{RAchange}\eeq
  This estimate is finite for all $A$, and since $(\mid \ve{r}-\ve{r}'\mid)_{\mathrm{max}}\propto A^{1/3}$, decreases like $A^{-2/3}$ for large $A$.

The momentum space matrix element between general normalised plane wave states of  momenta  $\ve{k},\,\ve{k}'$ that  corresponds to $\hat{\mathcal{T}}_{0,0}(E+\imath \epsilon; \ve{r},\ve{r}')$ is 
\beq\hat{\mathcal{T}}_{0,0}(E+\imath \epsilon; \ve{k},\ve{k}')&&=\int d\ve{r}\int d\ve{r}'\frac{\exp(-\imath\ve{k}.\ve{r})}{(2\pi)^{3/2}}\tilde{\mathcal{T}}_{0,0}(E+\imath \epsilon; \ve{r},\ve{r}')\frac{\exp(\imath\ve{k}'.\ve{r}')}{(2\pi)^{3/2}}\eol&&=\la\la \Psi(0,\ve{x}=0) \mid[ a_{\ve{k}},V]_-a^\dagger_{\ve{k}'}  \mid -\ve{k}',\psi_0 \ra\ra\eol&&+\la\la \Psi(0,\ve{x}=0) \mid[ a_{\ve{k}},V]_-G(E+\imath \epsilon)[V,a^\dagger_{\ve{k}'}]_-  \mid -\ve{k}',\psi_0 \ra\ra.\label{Tkk'}\eeq 

The quantities $\hat{\mathcal{T}}_{0,0}(E+\imath \epsilon; \ve{r},\ve{r}')$ and $\hat{\mathcal{T}}_{0,0}(E+\imath \epsilon; \ve{k},\ve{k}')$ can be thought of as matrix elements of an operator in $B$-space, $\hat{\mathcal{T}}_{0,0}(E+\imath \epsilon)$.  Using this identification, eq.(\ref{source2}) can know be expressed entirely in terms of operators and kets in $B$-space.
\beq(\tilde{\epsilon}_{k_0}+\imath \epsilon-\hat{T})\mid \xi_{E,\ve{k}_0}^\epsilon \ra&=&\hat{\mathcal{T}}_{0,0}(E+\imath \epsilon)(2\pi)^{3/2}\mid \ve{k}_0\ra+\imath \epsilon  (2\pi)^{3/2}\mid \ve{k}_0\ra, \eol &&\label{source6}\eeq
where 
\beq \la \ve{r}\mid \hat{T}\mid \ve{r}'\ra=-\frac{\hbar^2}{2\mu_{mA}}(\nabla^2_{\ve{r}}\delta{(\ve{r}-\ve{r}'})),\label{Tr2}\eeq
is the kinetic operator  in $B$-space for a particle of mass $\mu_{mA}$. The notation $\mid \,\,>$ is used for kets in $B$-space.

Eq.(\ref{source6}) has the  solution
 \beq \mid \xi_{E,\ve{k}_0}^\epsilon\ra&=& \hat{g}_0(E+\imath \epsilon)\hat{\mathcal{T}}_{0,0}(2\pi)^{3/2}\mid \ve{k}_0\ra+\imath \epsilon  (2\pi)^{3/2} \hat{g}_0(E+\imath \epsilon)\mid \ve{k}_0\ra\eol
 &=&\hat{g}_0(E+\imath \epsilon)\hat{\mathcal{T}}_{0,0}(2\pi)^{3/2}\mid \ve{k}_0\ra+\imath  (2\pi)^{3/2} \mid \ve{k}_0\ra, \eol &&\label{source7}\eeq
 where 
 \beq \hat{g}_0(E+\imath \epsilon)=\frac{1}{(E+\imath \epsilon-\hat{T})}.\label{g}\eeq
and where, for $E=\hbar k_0^2/2\mu_{mA}$, the result $\imath \epsilon   \hat{g}_0(E+\imath \epsilon)\mid \ve{k}_0)\ra=\mid \ve{k}_0\ra$  has been used.

 Hence
\beq \mid\xi_{E,\ve{k}_0}^\epsilon\ra
 &=&\hat{\Omega}(E+\imath \epsilon)(2\pi)^{3/2}\mid \ve{k}_0\ra, \eol &&\label{source8}\eeq
where
\beq \hat{\Omega}(E+\imath \epsilon)=(1+\hat{g}_0(E+\imath \epsilon)\hat{\mathcal{T}}_{0,0}). \label{omegaopt}\eeq

The notation used for $\hat{\mathcal{T}}_{0,0}(E+\imath \epsilon)$ reflects the fact that in the limit $\epsilon \rightarrow 0$,  between plane wave states of unit amplitude and on-energy-shell wave numbers, this operator gives the elastic scattering transition amplitude  and is related to the elastic scattering amplitude $f_{0,0}(\ve{k}'_0,\ve{k}_0)$. \beq \la \ve{k}'_0,\psi_0\mid T(E)\mid \ve{k}_0,\psi_0\ra&&=\int d\ve{r}\int d\ve{r}'\exp(-\imath\ve{k}'_0.\ve{r})\tilde{\mathcal{T}}_{0,0}(E+\imath \epsilon; \ve{r},\ve{r}')\exp(\imath\ve{k}_0.\ve{r}')\eol&&=\la \ve{k}'_0\mid V^\mathrm{opt}\mid \xi_{E,\ve{k}_0}^\epsilon \ra\eol&&=-2\pi\frac{\hbar^2}{\mu_{A}}f_{0,0}(\ve{k}'_0,\ve{k}_0).\label{fTelastic}\eeq 
 \subsection{Definition of the optical potential operator.}\label{optdef}

Standard many-body approaches to the definition of the optical model operator proceed through through the definition of the mass operator associated with the nucleon single-particle  Green's function\cite{DickhoffCharity2018}. A different path via the off-shell elastic transition operator is followed here.  The optical potential will be defined as the non-local operator in $B$-space that is related to $\hat{\mathcal{T}}_{0,0}(E+\imath \epsilon)$ by
  \beq \hat{V}^\mathrm{opt}(E+\imath \epsilon)=\hat{\mathcal{T}}_{0,0}(E+\imath \epsilon)-\hat{V}^\mathrm{opt}(E+\imath \epsilon)\hat{g}_0(E+\imath \epsilon) \hat{\mathcal{T}}_{0,0}(E+\imath \epsilon) ,\label{Vopt}\eeq
 
 This definition assumes that the operator $\hat{\Omega}$ defined in eq.(\ref{omegaopt}) has an inverse so that eq.(\ref{Vopt}) has the solution
  \beq \hat{V}^\mathrm{opt}(E+\imath \epsilon)=\hat{\mathcal{T}}_{0,0}(E+\imath \epsilon)\hat{\Omega}(E+\imath \epsilon)^{-1}.\label{Vopt2}\eeq
 To verify that the definition (\ref{Vopt}) produces an equation for  $ \xi_{E,\ve{k}_0}^\epsilon(\ve{r})$ of the form (\ref{source3}) note that  if $\hat{V}^\mathrm{opt}$ satisfies eq.(\ref{Vopt}) then
\beq \hat{V}^\mathrm{opt}(E+\imath \epsilon) \mid\xi_{E,\ve{k}_0}^\epsilon\ra
 &=&\hat{V}^\mathrm{opt}(E+\imath \epsilon)\hat{\Omega}(E+\imath \epsilon)(2\pi)^{3/2}\mid \ve{k}_0\ra, \eol&=&\hat{\mathcal{T}}_{0,0}(E+\imath \epsilon)(2\pi)^{3/2}\mid \ve{k}_0\ra.\label{source9}\eeq
   Referring to eq.(\ref{source6}), the equality (\ref{source9}) implies that $\xi_{E,\ve{k}_0}^\epsilon(\ve{r})$ satisfies 
\beq(E+\imath \epsilon-\hat{T}) \xi_{E,\ve{k}_0}^\epsilon(\ve{r})&=&\int d\ve{r}'\hat{V}^\mathrm{opt}(E+\imath \epsilon;\ve{r},\ve{r}') \xi_{E,\ve{k}_0}^\epsilon(\ve{r}')+\imath \epsilon  \exp(\imath\ve{k}_0.\ve{r}), \label{source10}\eeq
confirming that $\hat{V}^\mathrm{opt}$ does indeed plays the role of an optical potential operator.
 \subsection{Non-uniqueness of $\hat{V}^\mathrm{opt}$.}\label{nonuniqueVopt}
The optical potential $\hat{V}^\mathrm{opt}$ as defined by eq.(\ref{Vopt}) is not unique because the operator $\hat{\mathcal{T}}_{0,0}(E+\imath \epsilon)$ is not unique. Any operator $\hat{\mathcal{T}}'_{0,0}(E+\imath \epsilon)$ with the property
\beq \hat{\mathcal{T}}'_{0,0}(E+\imath \epsilon)\mid \ve{k}_0\ra =\hat{\mathcal{T}}_{0,0}(E+\imath \epsilon)\mid \ve{k}_0\ra,\label{T'k0}\eeq
when the half-on-shell condition $E=\hbar k_0^2/2\mu_{mA}$ is satisfied, will generate the same optical model wave function as $\hat{\mathcal{T}}_{0,0}(E+\imath \epsilon)$ when used in eq.(\ref{source3}). However, when used in eq.(\ref{Vopt}) it will give a different $\hat{V}^\mathrm{opt}(E)$ if its half-off-shell  momentum matrix elements $\hat{\mathcal{T}}'_{0,0}(E+\imath \epsilon)\mid \ve{k}\ra$ with $k \neq k_0 $ differ from those of $\hat{\mathcal{T}}_{0,0}(E+\imath \epsilon)$ because these matrix elements will contribute to the explicit solution, eq.(\ref{Vopt2}). An example of alternative definition of the optical potential is discussed in Section \ref{Dyson}. 

The particular choice of off-shell extensions for  $\hat{\mathcal{T}}_{0,0}(E+\imath \epsilon)$ as defined in eqs.(\ref{TE00}) and (\ref{Tkk'}) have the feature that the resulting operator conserve angular momentum in $B$ space (see Appendix \ref{rotinv}) and have standard transformation properties under time reversal (see Appendix  \ref{TRprops}) . 

\subsection{Interpretation of the optical potential operator defined by eq.(\ref{Vopt}). }\label{physinterp}
The formal expressions (\ref{TE00}) and (\ref{Tkk'}) for the the transition operator $\hat{\mathcal{T}}_{0,0}$ describe the many different reaction mechanisms that contribute to elastic scattering, such as direct and exchange scattering and heavy particle stripping. The relative importance of these mechanism varies with, for example, the incident energy and the mass of the target. In this section it is shown how some of these effects can be separated out from the formal expressions.

The definition in eq.(\ref{Tkk'}) is repeated here for convenience.
\beq\hat{\mathcal{T}}_{0,0}(E+\imath \epsilon; \ve{k},\ve{k}')&&=\la\la \Psi(0,\ve{x}=0) \mid[ a_{\ve{k}},V]_-a^\dagger_{\ve{k}'}  \mid -\ve{k}',\psi_0 \ra\ra\eol&&+\la\la \Psi(0,\ve{x}=0) \mid[ a_{\ve{k}},V]_-G(E+\imath \epsilon)[V,a^\dagger_{\ve{k}'}]_-  \mid -\ve{k}',\psi_0 \ra\ra.\label{Tkk'1}\eeq 
The first, Born, term on the right of eq.(\ref{Tkk'1}) is the sum of two terms that describe quite different physical processes. Using
\beq [ a_{\ve{k}},V]_-a^\dagger_{\ve{k}'}=\{[ a_{\ve{k}},V]_-,a^\dagger_{\ve{k}'}\}_+  -a^\dagger_{\ve{k}'}  [a_{\ve{k}},V]_-,\label{commantcom}\eeq
gives
\beq\hat{\mathcal{T}}^{\mathrm{Born}}_{0,0}=\tilde{\mathcal{T}}^{\mathrm{Born}(0)}_{0,0}+\tilde{\mathcal{T}}^{\mathrm {Born}(HPS)}_{0,0}\label{TBT1T2}\eeq
 where
\beq\hat{\mathcal{T}}^{\mathrm{Born}(0)}_{0,0}(E+\imath \epsilon; \ve{k},\ve{k}')=\la\la \Psi(0,\ve{x}=0) \mid\{[ a_{\ve{k}},V]_-,a^\dagger_{\ve{k}'}\}+  \mid -\ve{k}',\psi_0 \ra\ra,\label{TBorn1}\eeq
and
\beq\hat{\mathcal{T}}^{\mathrm{Born}(HPS)}_{0,0}(E+\imath \epsilon; \ve{k},\ve{k}')=-\la\la \Psi(0,\ve{x}=0) \mid a^\dagger_{\ve{k}'}[ a_{\ve{k}},V]_-  \mid -\ve{k}',\psi_0 \ra\ra.\label{TBorn2}\eeq
The significance of this separation becomes clear when the commutator and anticommutator are evaluated for a general additive, non-local two-body interaction $V$. In a momentum-space basis
\beq V=\frac{1}{4}\int d\ve{k}_1d\ve{k}_2d\ve{k}_3d\ve{k}_4\la\ve{k}_1,\ve{k}_2\mid V_{\mathcal{A}}\mid \ve{k}_3,\ve{k}_4\ra a^\dagger_{\ve{k}_1}a^\dagger_{\ve{k}_2}a_{\ve{k}_4}a_{\ve{k}_3},\label{Vgen2} \eeq
where the subscript $\mathcal{A}$ indicates that the matrix element involves normalised antisymmetrised two body states.

For this $V$ the commutator and anti-commutator that appear in the Born terms give 
\beq
  [a_{\ve{k}},V]_-=\frac{1}{2}\int d\ve{k}_2d\ve{k}_3d\ve{k}_4\la\ve{k},\ve{k}_2\mid V_{\mathcal{A}}\mid \ve{k}_3,\ve{k}_4\ra a^\dagger_{\ve{k}_2}a_{\ve{k}_4}a_{\ve{k}_3}, 
   \label{Vcommidentity22}\eeq
and
\beq
  \{[a_{\ve{k}},V]_-,a^\dagger_{\ve{k}'}\}_+=\int d\ve{k}_2d\ve{k}_4\la\ve{k},\ve{k}_2\mid V_{\mathcal{A}}\mid \ve{k}',\ve{k}_4\ra a^\dagger_{\ve{k}_2}a_{\ve{k}_4} ,\label{Vcommidentity32}\eeq
  or, in terms of non-antisymmetrised matrix elements of $V$
  \beq
  \{[a_{\ve{k}},V]_-,a^\dagger_{\ve{k}'}\}_+=\int d\ve{k}_2d\ve{k}_4\la\ve{k},\ve{k}_2\mid V (\mid  \ve{k}',\ve{k}_4\ra-\mid  \ve{k}_4,\ve{k}'\ra) a^\dagger_{\ve{k}_2}a_{\ve{k}_4} .\label{Vcommidentity322}\eeq

The anti-commutator term  (\ref{Vcommidentity32}) has the form of a nucleon-number conserving effective one-body operator including direct and exchange matrix elements of the nucleon-nucleon interaction. Scattering and excitations of the target  induced by this operator can proceed through low nucleon momentum components in the target ground state and  momentum transfer components that can be found in the short range interaction $V$. 

 On the other hand $ \hat{T}_{0,0}^{\mathrm{Born(HPS)}} $, eq.(\ref{TBorn2}), describes quite different processes. The explicit expression (\ref{Vcommidentity22})  shows that the state $[a_{\ve{k}},V]\mid -\ve{k}',\psi_0 \ra\ra$ that appears  in  eq.(\ref{TBorn2})  only involves interactions between pairs of nucleons that are initially in the target ground-state and produce an out-going nucleon with momentum $\ve{k}$. The  bra $\la\la \Psi(0,\ve{x}=0) \mid a^\dagger_{\ve{k}'}$ shows that the incoming momentum $\ve{k}'$ has to be found in the target ground state wavefunction. These are "heavy-particle stripping'' terms, in standard text-book language \cite{Austern}, page 93. In contrast, the first interactions in $ \hat{T}_{0,0}^{\mathrm{Born(0)}} $ are direct and exchange scattering of the incident nucleon by a target nucleon. It would be expected that, except at very low incident energies and/or very small $A$ when momentum components arising from target recoil may be comparable to the incident momentum, elastic scattering would be dominated by the processes described in $ \hat{T}_{0,0}^{\mathrm{Born(0)}} $ and their iterations.
 
For given values of $\ve{k}$ and $\ve{k}'$ the right-hand-side of eq.(\ref{TBorn1})  is the  matrix element of a  one-body interaction in Fock-space, $ \mathcal{V}$.
\beq\hat{\mathcal{T}}^{\mathrm{Born}(0)}_{0,0}(E+\imath \epsilon; \ve{k},\ve{k}')&&=\la\la \Psi(0,\ve{x}=0) \mid \mathcal{V}(\ve{k}, \ve{k}')\mid -\ve{k}',\psi_0 \ra\ra, \label{TBorn122}\eeq
where
\beq \mathcal{V}(\ve{k}, \ve{k}')&&=
\int d\ve{k}_2\int d\ve{k}_4\la\ve{k},\ve{k}_2\mid V_{\mathcal{A}}\mid \ve{k}',\ve{k}_4\ra a^\dagger_{\ve{k}_2}a_{\ve{k}_4}.
\label{Vkk'1}\eeq
In the next several subsections it will be shown that the complete transition operator $\tilde{\mathcal{T}}_{0,0}$ can be written 
\beq\hat{\mathcal{T}}_{0,0}=\tilde{\mathcal{T}}^{(1)}_{0,0}+\tilde{\mathcal{T}}^{\mathrm{Born}(HPS)}_{0,0}, \label{T1T2def0}\eeq
where $\tilde{\mathcal{T}}^{(1)}_{0,0}$, and not only the Born contribution $\tilde{\mathcal{T}}^{\mathrm{Born}(0)}_{0,0}$, can be expressed entirely in terms of matrix elements of  $\mathcal{V}$ in the $A$-nucleon sub-space.
\subsection{ Development of $\tilde{\mathcal{T}}^{(1)}_{0,0}$.} \label{T100}
The second term on the right in eq.(\ref{Tkk'1}) for $\hat{\mathcal{T}}_{0,0}$ can be expressed in terms of $ \mathcal{V}$, eq.(\ref{Vkk'1}), by examining the explicit formulae for the commutators $ [a_{\ve{k}},V]_-$ (see eq.(\ref{Vcommidentity22})) and $ [V,a^\dagger_{\ve{k}'},]_-$:
\beq
  [a_{\ve{k}},V]_-&&=\frac{1}{2} \int d\ve{k}_3\mathcal{V}(\ve{k}, \ve{k}_3) a_{\ve{k}_3}. 
   \label{Vcommidentity222}\eeq
   and
 \beq [V,a^\dagger_{\ve{k}'}]_-&&=\frac{1}{2}\int d\ve{k}_3 a^\dagger_{\ve{k}_3}  \mathcal{V}(\ve{k}_3, \ve{k}'). 
   \label{Vcommidentity223}\eeq
   
Note that if the nucleon-nucleon interaction  $V$ is Hermitian  then
\beq (\mathcal{V}(\ve{k}', \ve{k}))^\dagger=\mathcal{V}(\ve{k}, \ve{k}'). \label{VHemiticity}\eeq

 Eq.(\ref{Tkk'1}) for $\hat{\mathcal{T}}_{0,0}$ can  now be  written
\beq\hat{\mathcal{T}}_{0,0}=\tilde{\mathcal{T}}^{(1)}_{0,0}+\tilde{\mathcal{T}}^{\mathrm{Born}(HPS)}_{0,0}, \label{T1T2def}\eeq
where
\beq\hat{\mathcal{T}}^{(1)}_{0,0}(E+\imath \epsilon; \ve{k},\ve{k}')&&=\la\la \Psi(0,\ve{x}=0) \mid \mathcal{V}(\ve{k}, \ve{k}') \mid -\ve{k}',\psi_0 \ra\ra\eol&&+\frac{1}{4}\int d\ve{k}'_3\int d\ve{k}'_3\la\la \Psi(0,\ve{x}=0) \mid\mathcal{V}(\ve{k}, \ve{k}_3) a_{\ve{k}_3}G(E+\imath \epsilon) a^\dagger_{\ve{k}'_3}\mathcal{V}(\ve{k}'_3, \ve{k}')  \mid -\ve{k}',\psi_0 \ra\ra.\label{Tkk'3}\eeq
It will be shown below that, acting on the $A$-nucleon intermediate states that contribute to the Green's function term in eq.(\ref{Tkk'3}), the operator  $G(E+\imath \epsilon) a^\dagger_{\ve{k}'_3}$ can be expressed in terms of  matrix elements of $ \mathcal{V}$ in the $A$-nucleon sub-space.

 For a translationally invariant nucleon-nucleon interaction $V$ matrix elements of the operator  $\mathcal{V}$  satisfy 
   \beq \la\la -\ve{k}'',\psi_n\mid \mathcal{V}(\ve{k}, \ve{k}')\mid -\ve{k}''',\psi_m \ra\ra=(2\pi)^3\delta(\ve{k}-\ve{k}''-(\ve{k}'-\ve{k}'''))\la\la  \Psi(n,\ve{x}=0)\mid \mathcal{V}(\ve{k}, \ve{k}')
 \mid -\ve{k}''', \psi_m \ra\ra.
\label{V1memcon}\eeq
This result is most simply derived by noting that the delta-function dependence on the momenta  $\ve{k},\ve{k}'',\ve{k}',\ve{k}'''$ follows from the momentum conservation properties of the matrix elements of $V$ and the structure of $ \mathcal{V}$ in terms of creation and destruction operators. Choosing $\ve{k},\ve{k}',\ve{k}'''$ as the independent variables and integrating over $\ve{k}''$ using
  \beq\int d\ve{k}"\exp(\imath \ve{k}".\ve{x}) \mid -\ve{k}",\psi_n\ra\ra=(2\pi)^3\mid\Psi(n, \ve{x})\ra\ra.\label{kx}\eeq
gives the result (\ref{V1memcon}).
 
 A similar useful formula results when $\ve{k},\ve{k}',\ve{k}''$ are chosen as the independent variables:
   \beq \la\la -\ve{k}'',\psi_n\mid \mathcal{V}(\ve{k}, \ve{k}')\mid -\ve{k}''',\psi_m \ra\ra=(2\pi)^3\delta(\ve{k}-\ve{k}''-(\ve{k}'-\ve{k}'''))\la\la  -\ve{k}'', \psi_n\mid \mathcal{V}(\ve{k}, \ve{k}')
 \mid\Psi(m,\ve{x}=0) \ra\ra.
\label{V1memcon3}\eeq
Eqs.(\ref{V1memcon}) and (\ref{V1memcon3}) are special cases of a more general result derived in the Appendix \ref{matrixelementsOps}.

Note that eqs.(\ref{V1memcon}) and (\ref{V1memcon3}) imply that
\beq\la\la  -\ve{k}'', \psi_n\mid \mathcal{V}(\ve{k}, \ve{k}')
 \mid\Psi(m,\ve{x}=0) \ra\ra=\la\la  \Psi(n,\ve{x}=0)\mid \mathcal{V}(\ve{k}, \ve{k}')
 \mid -(\ve{k}''+\ve{k}'-\ve{k}), \psi_m \ra\ra. \label{MEeq}\eeq
 
 Using eqs.(\ref{V1memcon}) and (\ref{V1memcon3})  together with the result, valid for a translationally invariant  $H$,
\beq \la\la-\ve{k}_3''',\psi_n\mid a_{\ve{k}_3''}G(E+\imath \epsilon)a^\dagger_{\ve{k}'_3}\mid -\ve{k}'_3,\psi_{n'}\ra\ra=\delta(\ve{k}_3'''-\ve{k}_3'')(2\pi)^3\la\la\Psi(n, \ve{x}=0)\mid a_{\ve{k}_3''}G(E+\imath \epsilon)a^\dagger_{\ve{k}'_3}\mid  -\ve{k}'_3,\psi_{n'}\ra\ra,\label{kGk'}\eeq
allows eq.(\ref{Tkk'3}) to be written
\beq\tilde{\mathcal{T}}^{(1)}_{0,0}(E+\imath \epsilon; \ve{k},\ve{k}')&&=\la\la\Psi(0,\ve{x}=0)\mid\mathcal{V}(\ve{k}, \ve{k}')\mid-\ve{k}',\psi_0 \ra\ra\eol &&+\frac{1}{4}\sum_{n,n'}\int d\ve{k}'_3\int d\ve{k}''_3\la\la\Psi(0,\ve{x}=0)\mid\mathcal{V}(\ve{k}, \ve{k}_3'') \mid-\ve{k}''_3,\psi_n\ra\ra\la\la\Psi(n, \ve{x}=0)\mid a_{\ve{k}_3''}G(E+\imath \epsilon)a^\dagger_{\ve{k}'_3}\mid -\ve{k}'_3,\psi_{n'}\ra\ra\eol&&\times\la\la\Psi(n',\ve{x}=0) \mid \mathcal{V}(\ve{k}_3', \ve{k}') \mid-\ve{k}',\psi_0 \ra\ra .\eol&&\label{Tkk'4}\eeq
It is convenient here to generalise the concept of $B$-space introduced at the beginning of Section \ref{source} to  include the space of $A$-nucleon target states $\psi_n$. These target nucleons interact with the fictitious particle of reduced mass introduced in Section \ref{source} but they are not identical to it. A basis in this space is, by definition, an orthonormal set of states $\mid \ve{k},n \ra, n=0,\dots,\infty,$ where the fictitious particle has momentum $\ve{k}$ and the target nucleons have an intrinsic state $n$ and a total momentum $-\ve{k}$.  The interaction between the particle and the target is described by an operator $\hat{\mathcal{U}}$ in this extended space with matrix elements defined in terms of the Fock-space elements of $\bar{\mathcal{V}}$  by
\beq \la\ve{k},n  \mid\hat{ \mathcal{U}}\mid \ve{k}',n'  \ra=\frac{1}{2}\la\la\Psi(n,\ve{x}=0) \mid\mathcal{V}(\ve{k}, \ve{k}') \mid-\ve{k}',\psi_{n'}\ra\ra,\label{Uint}\eeq
where $\mathcal{V}$ is the interaction defined in eq.(\ref{Vkk'1}). The Hamiltonian of the system "fictitious particle+$A$ target nucleons" is
\beq\hat{\mathcal{H}}=\hat{T}+\hat{h}_A+\hat{\mathcal{U}},\label{HpartA}\eeq
where $\hat{T}$ is the kinetic energy operator associated with a particle of mass $\frac{A}{(A+1)}m$ and $\hat{h}_A$ is diagonal in the $\mid \ve{k},n \ra$ basis with eigenvalues $E_n$.
\beq \la \ve{k}',n'\mid\hat{h}_A \mid \ve{k},n \ra =\delta(\ve{k}-\ve{k}')\delta_{n,n'}E_n.\label{hA}\eeq
The introduction  of this extended space is useful because  the Fock-space states $G(E+\imath \epsilon)a^\dagger_{\ve{k}}\mid -\ve{k},\psi_n\ra\ra$ that appear in eq.(\ref{Tkk'3}) can be expressed in terms of the Green's function associated with the  Hamiltonian $\hat{\mathcal{H}}$.

 It is shown in  Appendix \ref{Gkn},  that the states $G(E+\imath \epsilon)a^\dagger_{\ve{k}}\mid -\ve{k},\psi_n\ra\ra$, $E$ fixed $\ve{k},\,n,$ arbitrary, satisfy the set of coupled equations eq.(\ref{Villars155}). In terms of matrix elements in extended $B$-space these equations can be written
\beq  \int d{k}'\sum_{n'}G(E^+)a^{\dagger}_{\ve{k}'}\mid- \ve{k}', \psi_{n'} \ra\ra \la\ve{k}',n'\mid(E^+-\hat{\mathcal{H}})\mid \ve{k},n\ra=
a^{\dagger}_{\ve{k}}\mid- \ve{k}, \psi_n \ra\ra,
\label{GEkn}\eeq
 with solution 
\beq G(E+\imath \epsilon)a^{\dagger}_{\ve{k}}\mid- \ve{k}, \psi_{n} \ra\ra&&=\sum_{n'}\int d\ve{k}'
a^{\dagger}_{\ve{k}'}\mid- \ve{k}', \psi_{n'} \ra\ra \la\ve{k}',n'\mid \frac{1}{(E+\imath \epsilon-\hat{\mathcal{H}})}\mid \ve{k},n\ra.\label{GEkn2}\eeq
The quantity on the left is a ket in Fock space and the right-hand-side is a linear combination of Fock space kets with coefficients that are matrix elements in the extended $B$-space introduced before eq.(\ref{HpartA}). 

Using (\ref{GEkn2}) in eq.(\ref{Tkk'4}) for $\tilde{\mathcal{T}}^{(1)}_{0,0}$ gives
\beq\tilde{\mathcal{T}}^{(1)}_{0,0}(E+\imath \epsilon; \ve{k},\ve{k}')&&=\la\la\Psi(0,\ve{x}=0)\mid\mathcal{V}(\ve{k}, \ve{k}')\mid-\ve{k}',\psi_0 \ra\ra\eol &&+\frac{1}{4}\sum_{n,n'}\int d\ve{k}'_3\int d\ve{k}''_3\la\la\Psi(0,\ve{x}=0)\mid\mathcal{V}(\ve{k}, \ve{k}''_3) \mid-\ve{k}''_3,\psi_n\ra\ra\eol&&\times\sum_{n''}\int d\ve{k}'''_3
\la\la\Psi(n, \ve{x}=0)\mid a_{\ve{k}_3''}a^{\dagger}_{\ve{k}'''_3}\mid- \ve{k}'''_3, \psi_{n''} \ra\ra \la\ve{k}'''_3,n''\mid \frac{1}{(E+\imath \epsilon-\hat{T}-\hat{h}_A-\hat{\mathcal{U}})}\mid \ve{k}'_3,n'\ra\eol&&\times\la\la\Psi(n',\ve{x}=0) \mid \mathcal{V}(\ve{k}', \ve{k}'_3) \mid-\ve{k}',\psi_0 \ra\ra,\eol&&\label{Tkk'5}\eeq
or, expressed entirely in terms of matrix elements of operators in the extended $B$-space
\beq\tilde{\mathcal{T}}^{(1)}_{0,0}(E+\imath \epsilon; \ve{k},\ve{k}')&&=2\la\ve{k},0  \mid\hat{ \mathcal{U}}\mid \ve{k}',0  \ra+\la\ve{k},0  \mid\hat{ \mathcal{U}}\hat{\mathcal{K}}_A\hat{\mathcal{G}}(E+\imath \epsilon)\hat{ \mathcal{U}} \mid\ve{k}',0 \ra,\label{Tkk'6}\eeq
where
 \beq \hat{\mathcal{G}}(E+\imath \epsilon)= \frac{1}{(E+\imath \epsilon-\hat{T}-\hat{h}_A-\hat{\mathcal{U}})},\label{Ghat}\eeq
 and 
 \beq  \la \ve{k}, n\mid\hat{\mathcal{K}}_A\mid\ve{k}'_3,n''\ra=\la\la\Psi(n, \ve{x}=0)\mid a_{\ve{k}}a^{\dagger}_{\ve{k}'}\mid- \ve{k}', \psi_{n''} \ra\ra \label{Ktilde}\eeq
 is the matrix of one-nucleon transition density matrices of the $A$-nucleon system in momentum space.
 \subsection{Summary of Section \ref{opt model}.}\label{summ1}
 From eqs. (\ref{TBorn2}) and (\ref{Tkk'6}) the complete exact formula for the off-shell transition matrix is  
 \beq\tilde{\mathcal{T}}_{0,0}(E+\imath \epsilon; \ve{k},\ve{k}')&&=2\la\ve{k},0  \mid\hat{ \mathcal{U}}\mid \ve{k}',0  \ra+\la\ve{k},0  \mid\hat{ \mathcal{U}}\hat{\mathcal{K}}_A\hat{\mathcal{G}}(E+\imath \epsilon)\hat{ \mathcal{U}} \mid\ve{k}',0 \ra\ra\eol&& -\la\la \Psi(0,\ve{x}=0) \mid a^\dagger_{\ve{k}'}[ a_{\ve{k}},V]_-  \mid -\ve{k}',\psi_0 \ra\ra.\label{Texact}\eeq

 Explicit expressions relating the matrix elements of  $\hat{ \mathcal{U}}$ in the $\mid \ve{k},n\ra$ basis and  matrix elements of the nucleon-nucleon interaction $V$ between Fock space states follow from eqs.(\ref{Uint}) and (\ref{Vkk'1}):
   \beq \la\ve{k},n  \mid\hat{ \mathcal{U}}\mid \ve{k}',n'  \ra&&=\frac{1}{2}\la\la\Psi(n,\ve{x}=0) \mid\mathcal{V}(\ve{k}, \ve{k}') \mid-\ve{k}',\psi_{n'}\ra\ra \eol&&=\frac{1}{2}\int d\ve{k}_2\int d\ve{k}_4\la\ve{k},\ve{k}_2\mid V_{\mathcal{A}}\mid \ve{k}',\ve{k}_4\ra  \la\la\Psi(n,\ve{x}=0) \mid a^\dagger_{\ve{k}_2}a_{\ve{k}_4}\mid-\ve{k}',\psi_{n'}\ra\ra.\label{Uint2}\eeq
   The first two terms on the right in eq.(\ref{Texact}) are expressed in terms of matrix elements of a one-body nucleon-number-conserving nucleon-nucleus interaction between eigenstates of the $A$ nucleon Hamiltonian. In this sense they have a similar structure to Feshbach's theory of the optical model \cite{Fes58}, but with both direct and exchange components of the nucleon-nucleon interaction included in $\hat{ \mathcal{U}}$. However, because of the factor $2$ in front of the first term and the occurrence of $\hat{\mathcal{K}}_A$ the first two terms do not have the standard form expected for the transition matrix associated with the Hamiltonian $\hat{\mathcal{H}}$ of eq.(\ref{HpartA}). If they did, and the "heavy-particle stripping term" were neglected, the solution of eq,(\ref{Vopt}) for the nucleon optical model operator would simply be the operator defined by Feshbach (see Appendix (\ref{FeshbachTheory})) corresponding to $\hat{\mathcal{H}}$ but with a nucleon with the appropriate nucleon-target reduced mass and matrix elements of the nucleon-nucleon corrected for recoil as in eq.(\ref{Uint2}).  It will be shown in sub-Section (\ref{quasi free}) that in the weak-binding  limit \footnote{The terminology of \cite{goldberger} is used.} this identification can be made with a modified definition of $\hat{ \mathcal{U}}$, but there appears to be no general justification for optical model developments that attempt to take antisymmetry into account in Feshbach's approach by simply  adding knock-out exchange terms, even when the "heavy-particle stripping term" is neglected.
   \section{The heavy-particle stripping term and the Lehmann, Symanzik and Zimmermann representation.}\label{HPS term}
  In standard many-body theories of the optical model that make the link with the nucleon single-particle Green's function, the second term on the right in eq.(\ref{Texact}), referred to here as the "heavy particle stripping term", is transformed using the identity
    \beq \la\la\Psi(0,\ve{x}=0)\mid  a^{\dagger}_{\ve{k}'}[ a_{\ve{k}},V]_-  \mid -\ve{k}',\psi_0 \ra\ra=\la\la\Psi(0,\ve{x}=0)\mid[V,a^{\dagger}_{\ve{k}'}]\frac{1}{(E_0+\frac{\epsilon_{k}}{A}-\epsilon_{k'}-H)}[ a_{\ve{k}},V]_-  \mid -\ve{k}',\psi_0 \ra\ra.\label{LSZdev22}\eeq
 This result is derived in Appendix \ref{LSZid}. The convention of a zero value for the target ground state intrinsic energy, $E_0$, has been abandoned temporarily to aid comparison with other treatments.\cite{Villars1967},\cite{DickhoffCharity2018}.  Note that for a stable ground state the denominator $(\frac{\epsilon_{k}}{A}-\epsilon_{k'}+E_0-H)$ never vanishes for $A>1$.
 
 Eq.(\ref{LSZdev22}) displays the "heavy particle stripping " contribution in terms of coupling between the $A$-nucleon ground state and a  complete set of $(A-1)$-nucleon intermediate states.  
  Using eq.(\ref{LSZdev22}) and eqs.(\ref{Tkk'1})-(\ref{TBorn2}) the complete on-shell ($\epsilon_{k'}=\epsilon_{k}$) elastic transition amplitude can be written
   \beq\hat{\mathcal{T}}_{0,0}(E_0+\frac{(A+1)}{A}\epsilon_{k}; \ve{k},\ve{k}')&&=\la\la \Psi(0,\ve{x}=0) \mid\{[ a_{\ve{k}},V]_-,a^\dagger_{\ve{k}'}\}  \mid -\ve{k}',\psi_0 \ra\ra\eol&&+\la\la \Psi(0,\ve{x}=0) \mid[ a_{\ve{k}},V]_-G(E_0+\frac{(A+1)}{A}\epsilon_{k}+\imath \epsilon)[V,a^\dagger_{\ve{k}'}]_-  \mid -\ve{k}',\psi_0 \ra\ra\eol&&-\la\la\Psi(0,\ve{x}=0)\mid[V,a^{\dagger}_{\ve{k}'}]\frac{1}{(E_0-\frac{(A-1)}{A}\epsilon_{k}-H)}[ a_{\ve{k}},V]_-  \mid -\ve{k}',\psi_0 \ra\ra.\label{TLSZ}\eeq 
   The structure of the denominator in the third term on the right can be understood as follows. A state of energy $E_0+\epsilon_{k'}/A-\epsilon_{k}$ is obtained when a nucleon of momentum $\ve{k}$ is knocked out of an incident channel $A$-nucleon state that has momentum $-\ve{k}$ and energy $E_0+\epsilon_{k'}/A$. The appropriate on-shell intermediate $(A-1)$-nucleon energy that should appear in the denominator is therefore  $E_0+\epsilon_{k'}/A-\epsilon_{k}$. For on-shell elastic scattering ($\epsilon_{k'}=\epsilon_{k}$) this reduces to $E_0-\frac{(A-1)}{A}\epsilon_{k}$. 

Eq.(\ref{TLSZ}) is the  Lehmann,  Symanzik and Zimmermann $(LSZ)$ formula for the elastic transition operator, when the third contribution is referred to as the hole term.  The LSZ formalism usually appears in a time-dependent version of scattering theory. The development here closely follow the methods of ref.\cite{Villars1967} modified to incorporate the requirements of translational invariance along the lines of \cite{Johnson17}. The LSZ formula lends itself well to a systematic development in terms of all contributions from the nucleon-nucleon interaction $V$, including those in the ground state wave function. A new definition of the one particle Green function used in this approach, including recoil corrections not included in standard treatments, is described in Section \ref{Greenfunction} below. 

Using the techniques described in sub-Section (\ref{T100}) the "heavy particle stripping term'', as expressed in the LSZ form, eq.(\ref{LSZdev22}), can also be written in terms of in terms of matrix elements of the one-body nucleon-nucleus interaction $\mathcal{V}$, but between eigenstates of the $(A-1)$ nucleon Hamiltonian.
\section{The weak-binding  limit.}\label{quasi free} 
The weak-binding limit is discussed at length in \cite{goldberger}, pages 775-780.  The essential idea is that for sufficiently high incident momentum $\ve{k}'$ and a sufficiently weakly bound target it is improbable that a nucleon in the target nucleus will be found with momentum greater than $\ve{k}'$. In the present context these ideas can be exploited by using an alternative exact form for the commutator $ [a_{\ve{k}},V]_-$, eq.(\ref{Vcommidentity22}) :
\beq
  [V,a^\dagger_{\ve{k}'}]_-&&=
\int d\ve{k}_2d\ve{k}_3d\ve{k}_4\theta(k_3-k_2)\la\ve{k}_3,\ve{k}_2\mid V_{\mathcal{A}}\mid \ve{k}',\ve{k}_4\ra a^\dagger_{\ve{k}_3}a^\dagger_{\ve{k}_2}a_{\ve{k}_4}. 
   \label{VcommQF2}\eeq
The theta function $\theta(x)$ is defined by
\beq \theta(x)=1,\,\,x>0\eol
=0,\,\,x<0. \label{theta}\eeq
The theta function means that the integration is restricted to the region $| \ve{k}_3| >|\ve{k}_2$.
There is no restriction on the limits on any of the other eigenvalues (momentum direction, spin and iso-spin) involved in the definition of the single nucleon states $\mid \ve{k}\ra$.
 
Under weak-binding assumptions, the ket  $ [V,a^\dagger_{\ve{k}'}]_-\mid-\ve{k}',\psi_0\ra\ra$ that appears in the second term on the right in eq.(\ref{Tkk'1}), the expression (\ref{VcommQF2}) can be replaced by the approximate form
\beq  [V,a^\dagger_{\ve{k}'}]_-&&\approx 
\int d\ve{k}_3 a^\dagger_{\ve{k}_3}\int d\ve{k}_2d\ve{k}_4\theta(k_3-k_2)\theta(k'-k_4)\la\ve{k}_3,\ve{k}_2\mid V_{\mathcal{A}}\mid \ve{k}',\ve{k}_4\ra a^\dagger_{\ve{k}_2}a_{\ve{k}_4}. \label{VcommQF3}\eeq
Similarily eq.(\ref{Vcommidentity22}) can be replaced by 
 \beq
  [a_{\ve{k}},V]_-\approx \int d\ve{k}_2d\ve{k}_3d\ve{k}_4\theta(k-k_2)\theta(k_3-k_4)\la\ve{k},\ve{k}_2\mid V_{\mathcal{A}}\mid \ve{k}_3,\ve{k}_4\ra a^\dagger_{\ve{k}_2}a_{\ve{k}_4}a_{\ve{k}_3}. 
   \label{VcommQF4}\eeq
To this approximation both  $ [a_{\ve{k}},V]_-$ and $ [V,a^\dagger_{\ve{k}'}]_-$ can be written in terms of a new one-body operator  $\bar{\mathcal{V}}$ defined by
\beq \bar{\mathcal{V}}(\ve{k}, \ve{k}')&&=
\int d\ve{k}_2\int d\ve{k}_4\theta(k-k_2)\theta(k'-k_4)\la\ve{k},\ve{k}_2\mid V_{\mathcal{A}}\mid \ve{k}',\ve{k}_4\ra a^\dagger_{\ve{k}_2}a_{\ve{k}_4},
\label{Vbarkk'}\eeq
as
\beq  [V,a^\dagger_{\ve{k}'}]_-\approx 
\int d\ve{k}_3 a^\dagger_{\ve{k}_3}\bar{\mathcal{V}}(\ve{k}_3, \ve{k}'),\,\,\,
 [a_{\ve{k}},V]_- \approx \int d\ve{k}_3\bar{ \mathcal{V}}(\ve{k}, \ve{k}_3)a_{\ve{k}_3}. \label{VcommQF5}\eeq

   Similarily, from eqs.(\ref{TBorn1}) and (\ref{Vcommidentity32}), the Born term $\hat{\mathcal{T}}^{\mathrm{Born}(0)}_{0,0}$ can also be written in terms of $\bar{\mathcal{V}}$. 
  \beq\hat{\mathcal{T}}^{\mathrm{Born}(0)}_{0,0}(E+\imath \epsilon; \ve{k},\ve{k}')&&=\la\la \Psi(0,\ve{x}=0) \mid\{[ a_{\ve{k}},V]_-,a^\dagger_{\ve{k}'}\}_+  \mid -\ve{k}',\psi_0 \ra\ra\eol&&=
\int d\ve{k}_2 d\ve{k}_4(\theta(k-k_2)\theta(k'-k_4)+\theta(k_2-k)\theta(k_4-k')+\theta(k_2-k)\theta(k'-k_4)+\theta(k-k_2)\theta(k_4-k'))\eol&&\times \la\ve{k},\ve{k}_2\mid V_{\mathcal{A}}\mid \ve{k}',\ve{k}_4\ra \la\la \Psi(0,\ve{x}=0) \mid a^\dagger_{\ve{k}_2}a_{\ve{k}_4}  \mid -\ve{k}',\psi_0 \ra\ra, \eol
&&\approx\la\la \Psi(0,\ve{x}=0) \mid\bar{\mathcal{V}}(\ve{k}, \ve{k}')\mid -\ve{k}',\psi_0 \ra\ra,\label{TBorn12}\eeq 
where in the neglected terms in the last line at least one nucleon of an interacting pair that has a momentum to be found in the ground state and greater than $\ve{k}$ or $\ve{k}'$.  The same approximation eliminates the "heavy particle stripping" term.

Proceeding as in subsection (\ref{T100}), eq.(\ref{Texact}) is replaced by the the weak-binding elastic transition matrix
 \beq\tilde{\mathcal{T}}^{WB}_{0,0}(E+\imath \epsilon; \ve{k},\ve{k}')&&=\la\ve{k},0  \mid\hat{ \mathcal{U}}_{WB}\mid \ve{k}',0  \ra+\la\ve{k},0  \mid\hat{ \mathcal{U}}_{WB}\hat{\mathcal{G}}_{WB}(E+\imath \epsilon)\hat{ \mathcal{U}}_{WB} \mid\ve{k}',0 \ra\ra,\label{TQF}\eeq
 where
  \beq \hat{\mathcal{G}}_{WB}(E+\imath \epsilon)= \frac{1}{(E+\imath \epsilon-\hat{\mathcal{H}}_{WB})},\label{GhatQF}\eeq
 and
  \beq\hat{\mathcal{H}}_{WB}=\hat{T}+\hat{h}_A+\hat{\mathcal{U}}_{WB}.\label{HpartAQF}\eeq

It is consistent with the weak-binding assumptions to use the approximation
\beq  \la \ve{k}, n\mid\hat{\mathcal{K}}_A\mid\ve{k}'_3,n'\ra&&=\la\la\Psi(n, \ve{x}=0)\mid a_{\ve{k}}a^{\dagger}_{\ve{k}'}\mid- \ve{k}', \psi_{n'} \ra\ra \eol
&&=\la\la\Psi(n, \ve{x}=0)\mid \delta(\ve{k}-\ve{k}')\mid- \ve{k}', \psi_{n'} \ra\ra-\la\la\Psi(n, \ve{x}=0)\mid a^{\dagger}_{\ve{k}'}a_{\ve{k}}\mid- \ve{k}', \psi_{n''} \ra\ra\eol&&\approx  \delta(\ve{k}-\ve{k}')\delta_{n,n'}.\label{Ktilde2}\eeq
The interaction $\hat{ \mathcal{U}}_{WB}$ that appears in $\tilde{\mathcal{T}}^{WB}_{0,0}$ is an operator in extended $B$-space and is defined in terms of the nucleon-nucleon interaction $V$ by, \emph{c.f.} eq.(\ref{Uint2}),
 \beq \la\ve{k},n  \mid\hat{ \mathcal{U}}_{WB}\mid \ve{k}',n'  \ra\!\!\!\!\!\!\!&&=\la\la\Psi(n,\ve{x}=0) \mid\bar{\mathcal{V}}(\ve{k}, \ve{k}') \mid-\ve{k}',\psi_{n'}\ra\ra \eol&&=\int d\ve{k}_2\int d\ve{k}_4\theta(k-k_2)\theta(k'-k_4)\la\ve{k},\ve{k}_2\mid V_{\mathcal{A}}\mid \ve{k}',\ve{k}_4\ra  \la\la\Psi(n,\ve{x}=0) \mid a^\dagger_{\ve{k}_2}a_{\ve{k}_4}\mid-\ve{k}',\psi_{n'}\ra\ra.\label{UintQF}\eeq
Eq.(\ref{TQF}) will be recognised as the standard formula for the off-shell elastic transition matrix associated with the Hamiltonian (\ref{HpartAQF}), including all target excitations induced by the interaction $\hat{ \mathcal{U}}_{WB}$. This is just the physical system studied by Feshbach \cite{Fes58}, except that here all nucleon-nucleon interactions include both direct and exchange terms and with appropriate modifications of matrix elements to take into account recoil and the weak binding assumption.  It can immediately be deduced using the techniques outlined in Appendix \ref{FeshbachTheory} that an alternative expression for the operator in $B$-space, $\hat{T}^{WB}_{0,0}$, is
   \beq \hat{T}^{WB}_{0,0}=U_{0,0}^{WB\mathrm{opt}}+U_{0,0}^{WB\mathrm{opt}}\frac{1}{E+\imath \epsilon-\hat{T}-U_{0,0}^{WB\mathrm{opt}}}U_{0,0}^{WB\mathrm{opt}},
   \label{T001Feshbach} \eeq
   where, in the notation of eq.(\ref{UintQF}),
   \beq \la\ve{k} \mid U_{0,0}^{WB\mathrm{opt}}\mid \ve{k}'\ra=\la\ve{k},n=0  \mid \hat{U}_{WB\mathrm{opt}}\mid \ve{k}',n=0  \ra, \label{Uopt00}\eeq
   and $\hat{U}_{WB\mathrm{opt}}$ is defined as in eq.(\ref{TU8}) of Appendix \ref{FeshbachTheory} with $\hat{V}$ replaced by $\hat{ \mathcal{U}}_{WB}$.
This implies that according to the definition (\ref{Vopt}) of Section \ref{optdef}, in the weak-binding limit the optical potential operator is
\beq V^\mathrm{opt}= U_{0,0}^{WB\mathrm{opt}}.\label{VoptFesh} \eeq
In particular, if the incident energy is below the threshold for exciting the target the operator $\hat{U}_{WB\mathrm{opt}}$ is Hermitian and so is the predicted optical potential $V^\mathrm{opt}$. If all target excitations are neglected the corresponding optical potential is the ground state expectation value of the truncated nucleon-nucleon potential defined in $\hat{ \mathcal{U}}_{WB}$, eq.(\ref{UintQF}).

 It also follows from this analysis that in the weak-binding limit Watson's multiple scattering theory and the associated optical model can be  modified to include antisymmetry and translational invariance by simply replacing the nucleon-nucleon interaction by the antisymmetrised  and truncated form that appears in eq.(\ref{UintQF}). This result is consistent with the work of \cite{TW} on the weak binding limit and described in \cite{goldberger}, pages 775-780, but without recoil corrections. 
 \section{Nucleon single-particle Green's function and the Dyson self-energy.}  \label{GreenfunctionDyson}
 \subsection{The single-particle Green's function.} \label{Greenfunction}The off-shell elastic transition operator defined by eqs.(\ref{Tkk'1}), (\ref{commantcom}) and (\ref{LSZdev22}) has the form
\beq\hat{\mathcal{T}}_{0,0}(E+\imath\epsilon; \ve{k},\ve{k}')&&=\la\la \Psi(0,\ve{x}=0) \mid\{[ a_{\ve{k}},V]_-,a^\dagger_{\ve{k}'}\}  \mid -\ve{k}',\psi_0 \ra\ra\eol&&+\la\la \Psi(0,\ve{x}=0) \mid[ a_{\ve{k}},V]_-\frac{1}{(E+\imath\epsilon-H)}[V,a^\dagger_{\ve{k}'}]_-  \mid -\ve{k}',\psi_0 \ra\ra\eol&&+\la\la\Psi(0,\ve{x}=0)\mid[V,a^{\dagger}_{\ve{k}'}]\frac{1}{(\epsilon_{k'}-\frac{\epsilon_{k}}{A}+H)}[ a_{\ve{k}},V]_-  \mid -\ve{k}',\psi_0 \ra\ra.\label{TLSZoffshell}\eeq  
The requirements of translation invariance lead to the different energy parameters in the denominators of the two terms on the right. Correspondingly, the usual definition of the  nucleon single-particle Green's function that ignores translational invariance must be modified. The definition used here for general complex $\omega$ is
\beq
  G_{0,0}(\ve{k},\ve{k}'; \omega)=&& \!\!\!\!\!\!\!\la\la\Psi(0,\ve{x}=0)\mid  a _{\ve{k}}\frac{1}{\omega-H}a^\dagger _{\ve{k}'}\mid  -\ve{k}',  \psi_{0} \ra\ra\eol
  && \!\!\!\!\!\!\!+\la\la\Psi(0,\ve{x}=0)\mid  a ^\dagger_{\ve{k}'}\frac{1}{\omega-\frac{(\epsilon_{k'}+\epsilon_k)}{A}+H}a _{\ve{k}}\mid  -\ve{k'},  \psi_{0} \ra\ra. \eol &&\label{G1n}\eeq
   This differs from the usual definition, e.g., \cite{Villars1967}, eq.(3.65), page 321, by the $1/A$ terms in the energy denominator in the second term and the appearance of a  ground state with its c.m. localised at the origin of coordinates in the bra and a ground state of total momentum $-\ve{k}'$ in the ket.  With this choices shown in eq.(\ref{G1n})  the two energy denominators reduce to the appropriate values shown in the transition matrix formula eq.(\ref{TLSZ}) in the on-shell limit, $\omega=\frac{(A+1)}{A}\epsilon_k,\,\,\,\,\,\epsilon_{k'}=\epsilon_k$. It will be shown below that these differences are essential if the Green's function is to have the standard relation to the on-shell transition matrix,\cite{Villars1967}, eq.3.66 page 321. 
  \subsection{Relation between Green's function  and the transition operator.}\label{G00T00} 
  The derivation of the relation between this Green's function, the free Green's function for a particle of reduced mass $\frac{A}{(A+1)}m$, and the transition operator defined in a momentum basis by eq.(\ref{Tkk'}) uses similar techniques to \cite{Villars1967}. Frequent use is made of the relation (\ref{Villars1}), but with an abbreviated notation. The  integers $1$ and $2$ will replace $\ve{k}'$ and $\ve{k}$, respectively, and $J_1$ and $J^\dagger_1$ will denote
  \beq J_1&=&[a_1,V], \eol
  J^\dagger_1&=&[V,a^\dagger_1]. \label{J2def11}\eeq
 In this notation eq.(\ref{Villars1}) becomes
   \beq  Ha^{\dagger}_{1}&=&a^{\dagger}_{1}H+\epsilon_{1} a^{\dagger}_{1}+J^\dagger_1,\eol
   Ha_{1} &=&a_{1}H-\epsilon_{1} a_{1}-J_1.\label{Villars42212}\eeq

The second term in $G_{0,0}(2,1;\omega)$, eq.(\ref{G1n})  gives
\beq
  (\omega-\frac{(A+1)}{A}\epsilon_2)G^{(2)}_{0,0}(2,1; \omega)=&&\!\!\!\!\!\la\la\Psi(0,\ve{x}=0)\mid  a ^\dagger_{1}\frac{1}{\omega-\frac{(\epsilon_1+\epsilon_2)}{A}+H)} (\omega-\frac{(A+1)}{A}\epsilon_2)a _{2}\mid  -\ve{k}_1,  \psi_{0} \ra\ra\eol&&
 \!\!\!\!\!\!\! \!\!\!\!\!\!\!\!\!\!=\la\la\Psi(0,\ve{x}=0)\mid  a ^\dagger_{1}\frac{1}{\omega-\frac{(\epsilon_2+\epsilon_2)}{A}+H} (\omega-\frac{(\epsilon_1+\epsilon_2)}{A}+H-\frac{(A+1)}{A}\epsilon_2+\frac{(\epsilon_1+\epsilon_2)}{A}-H)a _{2}\mid  -\ve{k}_1,  \psi_{0} \ra\ra\eol&&
\!\!\!\!\!\!\!\!\!\!\!\!\!\!\!\!\!=\la\la\Psi(0,\ve{x}=0)\mid  a ^\dagger_{1}a _{2}\mid  -\ve{k}_1,  \psi_{0} \ra\ra \eol&&\!\!\!\!\!\!\!\!\!\!\!\!\!\!\!\!\!+  \la\la\Psi(0,\ve{x}=0)\mid  a ^\dagger_{1}\frac{1}{\omega-\frac{(\epsilon_1+\epsilon_2)}{A}+H} (-\frac{(A+1)}{A}\epsilon_2a_2-a _{2}H+\epsilon_2a_2+J_2+\frac{(\epsilon_1+\epsilon_2)}{A}a_2\mid  -\ve{k}_1,  \psi_{0} \ra\ra\eol&&\!\!\!\!\!\!\!\!\!\!\!\!\!\!\!\!\!=
\la\la\Psi(0,\ve{x}=0)\mid  a ^\dagger_{1}a _{2}\mid  -\ve{k}_1,  \psi_{0} \ra\ra +  \la\la\Psi(0,\ve{x}=0)\mid  a ^\dagger_{1}\frac{1}{\omega-\frac{(\epsilon_1+\epsilon_2)}{A}+H} J_2\mid  -\ve{k}_1,  \psi_{0} \ra\ra,  \label{omega2G21}\eeq 
where use has been made of  
\beq H \mid  -\ve{k}_1,  \psi_{0} \ra\ra=\frac{\hbar^2(-\ve{k}_1)^2}{2Am}\mid  -\ve{k}_1,  \psi_{0} \ra\ra
=\frac{1}{A}\epsilon_1\mid  -\ve{k}_1,  \psi_{0} \ra\ra.\label{Hk1012}\eeq
 Similarly
  \beq
 (\omega-\frac{(A+1)}{A}\epsilon_1) (\omega-\frac{(A+1}{A}\epsilon_2)G^{(2)}_{0,0}(2,1; \omega)=&&\!\!\!\!\!(\omega-\frac{(A+1)}{A}\epsilon_1)
\la\la\Psi(0,\ve{x}=0)\mid  a ^\dagger_{1}a _{2}\mid  -\ve{k}_1,  \psi_{0} \ra\ra \eol&&\!\!\!\!\!\!\!\!\!\!\!\!\!\!\!\!\!+  \la\la\Psi(0,\ve{x}=0)\mid  a ^\dagger_{1}((\omega-\frac{(\epsilon_1+\epsilon_2)}{A}+H)-\frac{(A+1)}{A}\epsilon_1+\frac{(\epsilon_1+\epsilon_2)}{A}-H)\eol&&\times\frac{1}{\omega-\frac{(\epsilon_1+\epsilon_2)}{A}+H} J_2\mid  -\ve{k}_1,  \psi_{0} \ra\ra\eol
&&\!\!\!\!\!\!\!\!\!\!\!\!\!\!\!\!\!=(\omega-\frac{(A+1)}{A}\epsilon_1)
\la\la\Psi(0,\ve{x}=0)\mid  a ^\dagger_{1}a _{2}\mid  -\ve{k}_1,  \psi_{0} \ra\ra \eol&&\!\!\!\!\!\!\!\!\!\!\!\!\!\!\!\!\!+  \la\la\Psi(0,\ve{x}=0)\mid  a ^\dagger_{1}J_2\mid  -\ve{k}_1,  \psi_{0} \ra\ra\eol&&\!\!\!\!\!\!\!\!\!\!\!\!\!\!\!\!\!+\la\la\Psi(0,\ve{x}=0)\mid  (-\frac{(A+1)}{A}\epsilon_1a ^\dagger_{1}+\frac{(\epsilon_1+\epsilon_2)}{A}a ^\dagger_{1}-Ha ^\dagger_{1}+\epsilon_1a ^\dagger_{1}+J_1^\dagger)\eol&&\!\!\!\!\!\!\!\!\!\!\!\!\!\!\!\!\!\times\frac{1}{\omega-\frac{(\epsilon_1+\epsilon_2)}{A}+H} J_2\mid  -\ve{k}_1,  \psi_{0} \ra\ra\eol&&\!\!\!\!\!\!\!\!\!\!\!\!\!\!\!\!\!=(\omega-\frac{(A+1)}{A}\epsilon_1)
\la\la\Psi(0,\ve{x}=0)\mid  a ^\dagger_{1}a _{2}\mid  -\ve{k}_1,  \psi_{0} \ra\ra \eol&&\!\!\!\!\!\!\!\!\!\!\!\!\!\!\!\!\!+  \la\la\Psi(0,\ve{x}=0)\mid  a ^\dagger_{1}J_2\mid  -\ve{k}_1,  \psi_{0} \ra\ra\eol&&\!\!\!\!\!\!\!\!\!\!\!\!\!\!\!\!\!+\la\la\Psi(0,\ve{x}=0)\mid  J ^\dagger_{1}\frac{1}{\omega-\frac{(\epsilon_1+\epsilon_2)}{A}+H}J_2\mid  -\ve{k}_1,  \psi_{0} \ra\ra,\label{omega2G23}\eeq 
where use has been made of 
\beq H \mid  \Psi(0,\ve{x}=0) \ra\ra=\frac{\hbar^2(\ve{P})^2}{2Am}\mid  \Psi(0,\ve{x}=0) \ra\ra,\label{Hk101}\eeq
 together with the fact that inside the matrix element in eq.(\ref{omega2G23}) the Fock-space momentum operator $\ve{P}$ has the eigenvalue $-\ve{k}_2$. Hence
  \beq
 (\omega-\frac{(A+1)}{A}\epsilon_1) (\omega-\frac{(A+1}{A}\epsilon_2)G^{(2)}_{0,0}(2,1; \omega)=&&(\omega-\frac{(A+1)}{A}\epsilon_1)
\la\la\Psi(0,\ve{x}=0)\mid  a ^\dagger_{1}a _{2}\mid  -\ve{k}_1,  \psi_{0} \ra\ra \eol&&\!\!\!\!\!\!\!\!\!\!\!\!\!\!\!\!\!+  \la\la\Psi(0,\ve{x}=0)\mid  a ^\dagger_{1}J_2\mid  -\ve{k}_1,  \psi_{0} \ra\ra\eol&&\!\!\!\!\!\!\!\!\!\!\!\!\!\!\!\!\!+\la\la\Psi(0,\ve{x}=0)\mid  J ^\dagger_{1}\frac{1}{\omega-\frac{(\epsilon_1+\epsilon_2)}{A}+H}J_2\mid  -\ve{k}_1,  \psi_{0} \ra\ra\eol&&\eol&&\label{omega2G242}\eeq
Using very similar analysis the first term in $G_{0,0}(2,1; \omega)$ gives 
  \beq(\omega-\frac{(A+1)}{A}\epsilon_1)(\omega-\frac{(A+1)}{A}\epsilon_2) G^{(1)}_{0,0}(2,1; \omega_1)&&  =(\omega-\frac{(A+1)}{A}\epsilon_1)\la\la\Psi(0,\ve{x}=0)\mid a _{2}a^\dagger_1\mid  -\ve{k}_1,  \psi_{0} \ra\ra\eol&&+\la\la\Psi(0,\ve{x}=0)\mid J _{2}a_1^\dagger\mid  -\ve{k}_1,  \psi_{0} \ra\ra \eol &&
 +\Psi(0,\ve{x}=0)\mid J _{2}\frac{1}{\omega-H}J_1^\dagger\mid  -\ve{k}_1,  \psi_{0} \ra\ra,\eol&& \label{omega2G1312}\eeq
Putting together the results (\ref{omega2G1312}) and  (\ref{omega2G242}) gives 
 \beq
 (\omega-\frac{(A+1)}{A}\epsilon_1) (\omega-\frac{(A+1}{A}\epsilon_2)G_{0,0}(2,1; \omega)&=&(\omega-\frac{(A+1)}{A}\epsilon_1)
\delta(\ve{k}_2-\ve{k}_1)+\hat{\mathcal{T}}'_{0,0}(\omega; 2,1),\eol&&\eol&&\label{omega2G233}\eeq
where $\hat{\mathcal{T}}'_{0,0}$ is defined by
\beq\hat{\mathcal{T}}'_{0,0}(\omega; 2,1)&&=\la\la \Psi(0,\ve{x}=0) \mid \{J_2,a^\dagger_1 \} \mid -\ve{k}',\psi_0 \ra\ra\eol&&+\Psi(0,\ve{x}=0)\mid J _{2}\frac{1}{\omega-H}J_1^\dagger\mid  -\ve{k}_1,  \psi_{0} \ra\ra\eol&& +  \la\la\Psi(0,\ve{x}=0)\mid  J ^\dagger_{1}\frac{1}{\omega-\frac{(\epsilon_1+\epsilon_2)}{A}+H} J_2\mid  -\ve{k}_1,  \psi_{0} \ra\ra.\label{T'00}\eeq 
This differs from the off-shell transition matrix defined by  eq.(\ref{TLSZoffshell}), which in the present notation reads
\beq\hat{\mathcal{T}}_{0,0}(\omega; \ve{k},\ve{k}')&&=\la\la \Psi(0,\ve{x}=0) \mid\{J_2,a^\dagger_1\}  \mid -\ve{k}_1,\psi_0 \ra\ra\eol&&+\la\la \Psi(0,\ve{x}=0) \mid J_2\frac{1}{(\omega-H)}J_1^\dagger  \mid -\ve{k}_1,\psi_0 \ra\ra\eol&&+\la\la\Psi(0,\ve{x}=0)\mid J_1^\dagger\frac{1}{(\epsilon_1-\frac{\epsilon_2}{A}+H)}J_2 \mid -\ve{k}_1,\psi_0 \ra\ra.\label{TLSZoffshell2}\eeq
Note the different denominators in the third terms on the right in eqs.(\ref{T'00}) and (\ref{TLSZoffshell2}). However, fully on-shell, when $\epsilon_1=\epsilon_2$ and $\omega=\frac{(A+1)}{A}\epsilon_1+\imath \epsilon$,
 \beq \hat{\mathcal{T}}'_{0,0}(\frac{(A+1)}{A}\epsilon_1+\imath \epsilon; \ve{k},\ve{k}') =\hat{\mathcal{T}}_{0,0}(\frac{(A+1)}{A}\epsilon_1+\imath \epsilon; \ve{k},\ve{k}'),\label{T'Tonshell}\eeq
 in the limit $\epsilon \rightarrow 0^+$. In addition, in the same limit, eq.(\ref{omega2G233}) gives
 \beq
 (\omega-\frac{(A+1)}{A}\epsilon_1)^2G_{0,0}(2,1; \omega)&\rightarrow&\hat{\mathcal{T}}_{0,0}(\frac{(A+1)}{A}\epsilon_1+\imath \epsilon; 2,1),\label{omega2G22}\eeq
 in agreement with the standard relation between $G_{0,0}$ and the on-shell elastic transition matrix given in \cite{Villars1967}, eq.3.66, page 321 and \cite{DickhoffCharity2018}, eq.(66), page 8.
 
Eq.(\ref{omega2G233}) can be written as a relation between operators in the sub-space of $B$-space with the target in its ground state as
\beq
\hat{G}_{0,0}(\omega)&=&\hat{g}_0(\omega)
+\hat{g}_0(\omega)\hat{\mathcal{T}}'_{0,0}(\omega)\hat{g}_0(\omega),\label{Gg0}\eeq
where $g_0$ is the Green's function for a free particle of mass $\frac{A}{(A+1)}m$ defined in eq.(\ref{g}).
\subsection{The Dyson self-energy and an alternative definition of the optical potential.}\label{Dyson}
In  \cite{Rotureau16}, Section II eqs.(5) and (6),  the optical potential operator  is defined (in the plane wave basis used here)  as the Dyson self-energy $\Sigma'$ through the relation  
\beq \Sigma'=\hat{g}_0^{-1}-\hat{G}_{0,0}^{-1}.\label{Voptalt}\eeq
In Section \ref{optdef} the optical model operator is defined in eq.(\ref{Vopt2}) in terms of the transition operator $\hat{\mathcal{T}}_{0,0}$. The connection between the optical potential defined through eq.(\ref{Voptalt}) and the transition operator  $\hat{\mathcal{T}}'_{0,0}$ follows from 
 \beq
 \hat{g}_0^{-1}-\hat{G}_{0,0}^{-1}&=&\hat{\mathcal{T}}'_{0,0}\hat{g}_0\hat{G}_{0,0}^{-1}\eol
\hat{g}_0\hat{G}_{0,0}^{-1}&=&(1+\hat{g}_0\hat{\mathcal{T}}'_{0,0})^{-1}.
\label{Gg02}\eeq
These can be deduced from eq.(\ref{Gg0}) and together with the definition (\ref{Voptalt}) give
\beq \Sigma'=\hat{\mathcal{T}}'_{0,0}(1+\hat{g}_0\hat{\mathcal{T}}'_{0,0})^{-1}.\label{Voptalt2}\eeq
This result means $\Sigma'$ is related to $\hat{\mathcal{T}}'_{0,0}$ by the same formula as the optical model operator defined in eq.(\ref{Vopt2}) of Section \ref{optdef} is related to $\hat{\mathcal{T}}_{0,0}$.

The Dyson self-energy $\Sigma'$ is defined here using a one-particle Green's function, eq.(\ref{G1n}) and a free Green's function $g_0$, both modified to take into account the requirements of translational invariance. The resulting optical potential differs from the quantity defined in eq.(\ref{Vopt2}) of Section \ref{optdef}  because  $\hat{\mathcal{T}}'_{0,0}\neq\hat{\mathcal{T}}_{0,0}$ off-shell. Note however that when $\omega=\frac{(A+1)}{A}\epsilon_1$, $\epsilon_2$ arbitrary, i.e., half on-shell,  the two transition matrices are equal and therefore the distorted waves generated by the two optical potentials will be identical according to eq.(\ref{source7}). Of course, both these optical potentials will differ from the calculations of \cite{Rotureau16} because the Green's function that appear in  eq.(\ref{Voptalt}) differ in two ways to those used in \cite{Rotureau16}:

(i) The $1/A$ factors that appear in the denominator in the second term in eq.(\ref{G1n}). 

(ii)The matrix elements defining $\hat{G}_{0,0}$ in  eq.(\ref{G1n}) and used to define an optical potential through eq.(\ref{Voptalt}) involve a mixed basis with a localised ground state in the bra and a state of definite total momentum in the ket.

 Expressions for the time-dependent one-particle Green's function equivalent to the definition (\ref{G1n}) can be found in Appendix \ref{G00t}. 
 \section{Discussion and conclusions.}\label{DisCons}
It has been shown how a nucleon optical model operator for an $A$-nucleon target can be consistently defined within a translation invariant, completely antisymmetrised many-body theory without reference to a mean field concept. The distorted wave generated by the optical model potential defined in this way satisfies a quasi one-body scattering equation for a particle with a mass equal to the nucleon-target reduced mass. The distorted wave incorporates other recoil-effects exactly within  a vector space referred as $B$-space  where the configuration space operator $\hat{\ve{r}}$ can be interpreted as the separation of the incident nucleon and the target centre-of-mass. The basis of the method is the definition of a specific off-shell extension of the many-body transition matrix. This is used to define the optical model operator as the solution of an integral equation in barycentric space. 
    
    The particular off-shell extension chosen is shown to satisfy rotational invariance  and to have properties under time reversal that  agree with standard conventions. It is also shown that when 'heavy particle stripping " is ignored the transition matrix can be expressed entirely in terms of matrix elements in the $A$-nucleon sub-space of a one-body interaction constructed from the nucleon-nucleon interaction with exchange.  Similarly, the 'heavy particle stripping" term can be expressed in terms of matrix elements of the nucleon-nucleon interaction in the $(A-1)$-nucleon subspace. 
    
    Because the method is based on the transition matrix it is straightforward to relate it to standard methods based on the one-nucleon $G$-matrix and the Dyson self energy. The modifications of the definition of the $G$ matrix necessitated by translational invariance result in an optical potential that differs from the one defined in Section \ref{opt model}, although the corresponding distorted waves are identical if a translationally invariant transition matrix  is used in both cases.
    
In the method described here any theory that generates a  calculation of the off-shell extension of the many-body elastic transition matrix defined in eq.(\ref{TE00}) or eq.(\ref{Tkk'}) leads to a corresponding optical model operator through eq.(\ref{Vopt}). Knowledge the off-shell elastic transition matrix alone is sufficient to calculate the optical model distorted wave through eq.(\ref{source6}) or eq.(\ref{source7}), and there is then no need to make the final step to calculate the optical potential. However, an important application of the nucleon optical model concept is to  few-body theories of composite particle reactions, \emph{e.g.}, the $A(d,p)B$ reaction, as a tool for nuclear structure studies. For recent reviews to theoretical  and experimental work see \cite{Johnson2014} and \cite{Wimmer2018} and references therein. For these developments knowledge of the non-local nucleon optical model operator itself is essential. 
\section{Acknowledgements.}\label{acks}
Support from the UK Science and Technology Facilities Council through the grant STFC ST/000051/1 is acknowledged.
 
\appendix
  \section{Symmetry properties of the optical potential operator.}\label{symmetry}
 In understanding the symmetry properties of $V^\mathrm{opt}$ it is important to distinguish between symmetry transformations in $B$-space and the corresponding transformations in Fock-space. To this end it is convenient to define the $B$-space operator  $\hat{\mathcal{T}}_{0,0}(E+\imath \epsilon)$ with matrix elements $\hat{\mathcal{T}}_{0,0}(E+\imath \epsilon; \ve{k},\ve{k}')$ through the relation 
\beq\hat{\mathcal{T}}_{0,0}(E+\imath \epsilon)&&=\int d\ve{k}\int d\ve{k}'\mid \ve{k}\ra\la \ve{k}'\mid \times\hat{\mathcal{T}}_{0,0}(E+\imath \epsilon; \ve{k},\ve{k}'). \label{TE00op1}\eeq
As already noted, operators in $B$-space are indicated with a "hat", unless other notation make this unnecessary.
\subsection{Rotation invariance.}\label{rotinv}
Consider the transform under rotation of eq.(\ref{TE00op1}) by the unitary operator for an arbitrary rotation $\hat{\mathcal{R}}$.
\beq\hat{\mathcal{R}}\hat{\mathcal{T}}_{0,0}(E+\imath \epsilon)\hat{\mathcal{R}}^{-1}&&=\int d\ve{k}\int d\ve{k}'\hat{\mathcal{R}}\mid \ve{k}\ra\la \ve{k}'\mid\hat{\mathcal{R}}^{-1}\times \hat{\mathcal{T}}_{0,0}(E+\imath \epsilon; \ve{k},\ve{k}')\eol &&= \int d\ve{k}\int d\ve{k}'\mid \mathcal{R}\ve{k}\ra\la \mathcal{R}\ve{k}'\mid
\times \hat{\mathcal{T}}_{0,0}(E+\imath \epsilon; \ve{k},\ve{k}').\label{TE00op2}\eeq
Changing the variables of integration to $\ve{k}''=\mathcal{R}\ve{k}$ and $\ve{k}'''=\mathcal{R}\ve{k}'$ gives
\beq\hat{\mathcal{R}}\hat{\mathcal{T}}_{0,0}(E+\imath \epsilon)\hat{\mathcal{R}}^{-1}&&= \int d\ve{k}''\int d\ve{k}'''\mid \ve{k}''\ra\la\ve{k}'''\mid
\times \hat{\mathcal{T}}_{0,0}(E+\imath \epsilon; \mathcal{R}^{-1}\ve{k}'',\mathcal{R}^{-1}\ve{k}''').\label{TE00op3}\eeq
The contribution from the second term on the right of eq.(\ref{Tkk'}) to $\hat{\mathcal{T}}_{0,0}(E+\imath \epsilon; \mathcal{R}^{-1}\ve{k}'',\mathcal{R}^{-1}\ve{k}''')$ is
\beq \hat{\mathcal{T}}_{0,0}(E+\imath \epsilon; \mathcal{R}^{-1}\ve{k}'',\mathcal{R}^{-1}\ve{k}''')_{(2)} &&=\la\la\Psi(0,\ve{x}=0)\mid[a_{\mathcal{R}^{-1}\ve{k}''},V]\frac{1}{(E+\imath \epsilon-H)}[V,a^\dagger_{\mathcal{R}^{-1}\ve{k}'''}] \mid-\mathcal{R}^{-1}\ve{k}''',\psi_0\ra\ra\eol
&&=\la\la\Psi(0,\ve{x}=0)\mid\mathcal{R}^{-1}[a_{\ve{k}''},V]\frac{1}{(E+\imath \epsilon-H)}[V,a^\dagger_{\ve{k}'''}]\mathcal{R}\mid-\mathcal{R}^{-1}\ve{k}''',\psi_0\ra\ra\eol&&=\la\la\Psi(0,\ve{x}=0)\mid [ a_{\ve{k}''},V]\frac{1}{(E+\imath \epsilon-H)}[V,a^\dagger_{\ve{k}'''}]\mid-\ve{k}'''\psi_0\ra\ra,\eol &&\label{TE00op4}\eeq
where now $\mathcal{R}$ means the rotation operator in Folk-space corresponding to $\hat{\mathcal{R}}$  and it is  assumed $H$ and $V$ are invariant under rotations. For simplicity, the ground state $\psi_0$ has been taken to have zero spin so that
\beq  \mathcal{R} \mid\Psi(0,\ve{x}=0)\ra\ra= \mid\Psi(0,\ve{x})=0\ra\ra. \label{RPsi0x}\eeq 
The other  term in eq.(\ref{Tkk'}) transform in the same way and eq.(\ref{TE00op2}) can  be written  
\beq\hat{\mathcal{R}}\hat{\mathcal{T}}_{0,0}(E+\imath \epsilon)\hat{\mathcal{R}}^{-1}=\hat{\mathcal{T}}_{0,0}(E+\imath \epsilon).\label{TE00op6}\eeq

It follows from eqs. (\ref{TE00op6}) and (\ref{Vopt2}) that $V^\mathrm{opt}$ has the analogous property:
\beq \hat{\mathcal{R}}\hat{V}^\mathrm{opt}\hat{\mathcal{R}}^{-1}=\hat{V}^\mathrm{opt}. \label{VoptRot} \eeq
The rotational invariance of the elastic scattering $T$-matrix $ \langle \ve{k}'_0,0\mid T(E) \mid \ve{k}_0,0 \rangle$ appearing in eq.(\ref{fTelastic}) also follows from eq.(\ref{TE00op6}):
\beq \la (\mathcal{R}\ve{k}'_0),\psi_0\mid T(E)\mid (\mathcal{R}\ve{k}_0),\psi_0\ra=\la \ve{k}'_0,\psi_0\mid T(E)\mid \ve{k}_0,\psi_0\ra.\label{fTelasticRot}\eeq 

\subsection{Time reversal properties of $ \hat{T}_{0,0}$.}\label{TRprops}
In applications it is convenient to work with operators that behave in a specific way under transformation by the anti-unitary time-reversal operator $\mathcal{K}$. The conventional transformation property for transition operators in $B$-space is  
\beq\hat{ \mathcal{K}}\hat{\mathcal{T}}_{0,0}(E+\imath \epsilon)\hat{\mathcal{K}}^{-1}= (\hat{\mathcal{T}}_{0,0}(E+\imath \epsilon))^\dagger. \label{TRcond}\eeq
The proof of the result (\ref{TRcond}) starts from the representation given in eq.(\ref{TE00op1}). Consider the second term on the right of eq.(\ref{Tkk'}) as an example.
\beq(\hat{\mathcal{T}}_{0,0}(E+\imath \epsilon))_{(2)}^\dagger\!\!\!\!\!\!&&=\int d\ve{k}\int d\ve{k}'\mid \ve{k}'\ra\la \ve{k}\mid (\la\la\Psi(0,\ve{x}=0)\mid[a_{\ve{k}},V]\frac{1}{(E+\imath \epsilon-H)}[V,a^\dagger_{\ve{k}'}] \mid-\ve{k}',\psi_0\ra\ra])^*\eol
&&=\int d\ve{k}\int d\ve{k}'\mid \ve{k}'\ra\la \ve{k}\mid \la\la-\ve{k}',\psi_0\mid([a_{\ve{k}},V]\frac{1}{(E+\imath \epsilon-H)}[V,a^\dagger_{\ve{k}'}]) ^\dagger\mid\Psi(0,\ve{x}=0)\ra\ra
\eol
&&=\int d\ve{k}\int d\ve{k}'\mid \ve{k}\ra\la \ve{k}'\mid \la\la-\ve{k},\psi_0\mid[a_{\ve{k}},V]\frac{1}{(E-\imath \epsilon-H)}[V,a^\dagger_{\ve{k}'}]) \mid\Psi(0,\ve{x}=0)\ra\ra,\eol&& \label{TE00opdagger}\eeq
where in the last line the integration/summation variables $\ve{k}, \ve{k}'$ have been interchanged.

On the other hand, the effect of anti-linear time reversal transform operator on this term is
\beq\hat{\mathcal{K}}\hat{\mathcal{T}}^\epsilon_{0,0}(E+\imath \epsilon)_{(2)}\hat{\mathcal{K}}^{-1}\!\!\!\!\!\!&&=\int d\ve{k}\int d\ve{k}'\mid -\ve{k}\ra\la -\ve{k}'\mid\eol&&\times (\la\la\Psi(0,\ve{x}=0)\mid[a_{\ve{k}},V]\frac{1}{(E+\imath \epsilon-H)}[V,a^\dagger_{\ve{k}'}] \mid-\ve{k}',\psi_0\ra\ra])^*, \label{TE00opTR1}\eeq
where the defining property of the time reversal operator, $\hat{\mathcal{K}}\mid \ve{k}\ra=\mid -\ve{k}\ra$, has been used.

Changing the variables of integration/summation from $-\ve{k}$ to $\ve{k}$,  $-\ve{k}'$ to $\ve{k}'$ , and using $ \mathcal{K}^{-1}a_{\ve{k}}\mathcal{K}= a_{-\ve{k}}$ gives
\beq\hat{\mathcal{K}}\hat{\mathcal{T}}^\epsilon_{0,0}(E+\imath \epsilon)_{(2)}\hat{\mathcal{K}}^{-1}&&=\int d\ve{k}\int d\ve{k}'\mid \ve{k}\ra\la \ve{k}'\mid\eol&&\times (\la\la\Psi(0,\ve{x}=0)\mid[a_{-\ve{k}},V]\frac{1}{(E+\imath \epsilon-H)}[V,a^\dagger_{-\ve{k}'}] \mid\ve{k}',\psi_0\ra\ra)^*\eol &&=\int d\ve{k}\int d\ve{k}'\mid \ve{k}\ra\la \ve{k}'\mid\eol&&\times (\la\la\Psi(0,\ve{x}=0)\mid (\hat{\mathcal{K}}^{-1}[a_{\ve{k}},V]\frac{1}{(E-\imath \epsilon-H)}[V,a^\dagger_{\ve{k}'}] \hat{\mathcal{K}})\mid\ve{k}',\psi_0\ra\ra)^*.\eol && \label{TE00opTR2}\eeq
where $ \mathcal{K}^{-1}\psi(\ve{r})\mathcal{K}= \psi(\mathcal{K}^{-1}\ve{r})$ has been used.

A general property of the matrix elements of an arbitrary linear operator $O$ and its time reverse transform $\mathcal{K}^{-1}O\mathcal{K} $ is
\beq \la\la a \mid O \mid b\ra \ra^*= \la\la a' \mid (\mathcal{K}^{-1}O \mathcal{K})\mid b'\ra\ra \label{Kprop}\eeq
where
\beq \mid a'\ra\ra= \mathcal{K}^{-1}\mid a \ra\ra,\,\,\,\mid b' \ra\ra=\mathcal{K}^{-1}\mid b\ra\ra. \label{a'b'}\eeq
Applying this to the Fock-space matrix element in eq.(\ref{TE00opTR2}) gives
\beq \la\la\Psi(0,\ve{x}=0)\mid (\hat{\mathcal{K}}^{-1}[a_{\ve{k}},V]\frac{1}{(E-\imath \epsilon-H)}[V,a^\dagger_{\ve{k}'}] \hat{\mathcal{K}})\mid\ve{k}',\psi_0\ra\ra])^*&&=\la\la\Psi(0,\ve{x}=0)\mid [a_{\ve{k}},V]\frac{1}{(E-\imath \epsilon-H)}[V,a^\dagger_{\ve{k}'}] \mid-\ve{k}',\psi_0\ra\ra.\eol&&\label{TE00opTR3}\eeq
The target ground state has been taken to have zero spin for simplicity. Arbitrary phases can then be chosen so that the ground state wave function $\psi_0$ is unchanged under the action of $\mathcal{K}$.

Applying the result (\ref{TE00opTR3}), eq.(\ref{TE00opTR2}) becomes
\beq\hat{\mathcal{K}}\hat{\mathcal{T}}^\epsilon_{0,0}(E+\imath \epsilon)_{(2)}\hat{\mathcal{K}}^{-1}&&=\int d\ve{k}\int d\ve{k}'\mid \ve{k}\ra\la \ve{k}'\mid\eol&&\times \la\la\Psi(0,\ve{x}=0)\mid [a_{\ve{k}},V]\frac{1}{(E-\imath \epsilon-H)}[V, a^\dagger_{\ve{k}'}] \mid-\ve{k}',\psi_0\ra\ra.,\eol && \label{TE00opTR4}\eeq
Using the techniques set out in Appendix \ref{matrixelementsOps} it follows that
\beq \la\la\Psi(0,\ve{x}=0)\mid [a_{\ve{k}},V]\frac{1}{(E-\imath \epsilon-H)}[V, a^\dagger_{\ve{k}'}] \mid-\ve{k}',\psi_0\ra\ra&=&\la\la-\ve{k},\psi_0\mid [a_{\ve{k}},V]\frac{1}{(E-\imath \epsilon-H)}[V, a^\dagger_{\ve{k}'}] \mid\Psi(0,\ve{x}=0)\ra\ra,\eol&&\label{MEequality}\eeq
and hence by comparison with eq.(\ref{TE00opdagger})\beq  \hat{\mathcal{K}}\hat{\mathcal{T}}^\epsilon_{0,0}(E+\imath \epsilon)_{(2)}\hat{\mathcal{K}}^{-1}=(\hat{\mathcal{T}}_{0,0}(E+\imath \epsilon))_{(2)}^\dagger.\label{TR3result}\eeq
 The proof of the analogous result for the first term of eq.(\ref{Tkk'}) uses similar techniques. The final result is
\beq  \hat{\mathcal{K}}\hat{\mathcal{T}}^\epsilon_{0,0}(E+\imath \epsilon)\hat{\mathcal{K}}^{-1}=(\hat{\mathcal{T}}_{0,0}(E+\imath \epsilon))^\dagger.\label{TRresult}\eeq
It follows from the definition of $\hat{V}^\mathrm{opt}(E+\imath \epsilon)$ in terms of $\hat{\mathcal{T}}^\epsilon_{0,0}(E+\imath \epsilon)$ given in eq.(\ref{Vopt}) that
\beq  \hat{\mathcal{K}}\hat{V}^\mathrm{opt}(E+\imath \epsilon)\hat{\mathcal{K}}^{-1}=(\hat{V}^\mathrm{opt}(E+\imath \epsilon))^\dagger.\label{TRVopt}\eeq
The implications of these results for the properties of distorted waves associated with $\hat{V}^\mathrm{opt}(E+\imath \epsilon)$ are described in the next subsection.

It should be noted that in the interests of conciseness important phases associated with the effect of the time reversal operator on spin eigenstates have been ignored in the derivation presented in this Appendix.

The implications of these results for the properties of distorted waves associated with $\hat{V}^\mathrm{opt}(E+\imath \epsilon)$ are described in the next subsection.
 \subsection{The scattering states $\mid \xi_{E,\ve{k}_0}^{(\pm)} \ra$ and time reversal requirements.}
In applications of the optical model to reaction models two different scattering states, $\mid \xi_{E,\ve{k}_0}^{(\pm)} \ra$,  associated with $\hat{V}^\mathrm{opt}$ are used. These are defined as the limit $\epsilon \rightarrow 0^+$ of $\mid \xi_{E,\ve{k}_0}^{\pm\epsilon} \ra$ where
\beq(E+\imath \epsilon-\hat{T}-\hat{V}^\mathrm{opt}(E+\imath \epsilon)) \mid \xi_{E,\ve{k}_0}^{+\epsilon} \ra&&=\imath \epsilon  (2\pi)^{3/2}\mid \ve{k}_0\ra,\eol
(E-\imath \epsilon-\hat{T}-(\hat{V}^\mathrm{opt}(E+\imath \epsilon))^\dagger) \mid \xi_{E,\ve{k}_0}^{-\epsilon} \ra&&=-\imath \epsilon  (2\pi)^{3/2}\mid \ve{k}_0\ra\label{xi+-}\eeq
The behaviour of the operators $\hat{\mathcal{T}}_{0,0}$ and $ \hat{V}_{\mathrm{LSZ}}^{\mathrm{opt},}$ under the time reversal transformation operator $\hat{\mathcal{K}}$ is discussed in Appendix \ref{TRprops}. Because (\ref{TRVopt}) is satisfied the two solutions $\mid \xi_{E,\ve{k}_0}^{\pm\epsilon}$ are related by
\beq \hat{\mathcal{K}} \mid \xi_{E,\ve{k}_0}^{(-)} \ra= \mid \xi_{E,-\ve{k}_0}^{(+)}\ra.\label{Kxi+-}\eeq
Note also that 
\beq(\hat{V}^\mathrm{opt}(E+\imath \epsilon))^\dagger&&=\hat{V}^\mathrm{opt}(E-\imath \epsilon),\eol
(\hat{\mathcal{T}}_{0,0}(E+\imath \epsilon))^\dagger&&=\hat{\mathcal{T}}_{0,0}(E-\imath \epsilon).\eol
\hat{\mathcal{K}}\mid \xi_{E,\ve{k}_0}^{-\epsilon} \ra&&=\mid \xi_{E,-\ve{k}_0}^{+\epsilon }\ra. \label{+-epsilon}\eeq

Another pair of solutions $\mid \tilde{\xi}_{E,\ve{k}_0}^{(\pm)} \ra$ are also needed when orthogonal sets of distorted waves corresponding to non-Hermitian optical potentials are required. These are defined as the limit $\epsilon \rightarrow 0^+$ of $\mid \tilde{\xi}_{E,\ve{k}_0}^{\pm\epsilon} \ra$ where
\beq(E+\imath \epsilon-\hat{T}-(\hat{V}^\mathrm{opt}(E+\imath \epsilon))^\dagger) \mid \tilde{\xi}_{E,\ve{k}_0}^{+\epsilon} \ra&&=\imath \epsilon  (2\pi)^{3/2}\mid \ve{k}_0\ra,\eol
(E-\imath \epsilon-\hat{T}-\hat{V}^\mathrm{opt}(E+\imath \epsilon) \mid \tilde{\xi}_{E,\ve{k}_0}^{-\epsilon} \ra&&=-\imath \epsilon  (2\pi)^{3/2}\mid \ve{k}_0\ra.\label{tildexi+-}\eeq
 The states $\mid \tilde{\xi}_{E,\ve{k}_0}^{(\pm)} \ra$  satisfy
\beq \la\tilde{\xi}_{E,\ve{k}_0}^{(+)} \mid\xi_{E,\ve{k}_0}^{(+)} \ra&&=(2\pi)^3\delta(\ve{k}'_0-\ve{k}_0),\eol
\la\tilde{\xi}_{E,\ve{k}_0}^{(-)} \mid\xi_{E,\ve{k}_0}^{(-)} \ra&&=(2\pi)^3\delta(\ve{k}'_0-\ve{k}_0)\label{tildexiorth} \eeq

\subsection{Matrix elements of a class of momentum conserving operators.}\label{matrixelementsOps}
This Appendix is concerned with  operators of the form
\beq  O_1a^{\dagger}_{\ve{k}_1}O_2a_{\ve{k}_2}O_3,\label{specOps}\eeq
where the $O_i$ are arbitrary momentum conserving operators in Fock space. The operators (\ref{specOps}) have a simple form in the  basis $\mid -\ve{k},\psi_n\ra\ra$ in which the $A$-nucleon intrinsic state $n$ has a total momentum $-\ve{k}$. These states are normalised so that 
\beq  \la\la-\ve{k}',\psi_{n'}\mid -\ve{k},\psi_n\ra\ra=(2\pi)^3\delta_{n',n}\delta(\ve{k}'-\ve{k}), \label{norm n k}\eeq
and they are related to states $\mid\Psi(n, \ve{x})\ra\ra$, in which the c.m. is located at $\ve{x}$ and defined in \cite{Johnson17}, by
\beq \mid -\ve{k},\psi_n\ra\ra=\int d\ve{x}\exp(-\imath \ve{k}.\ve{x})\mid\Psi(n, \ve{x})\ra\ra.\label{kx2}\eeq

In this basis
\beq \la\la -\ve{k}, \psi_{n} \mid  O_1a^{\dagger}_{\ve{k}_1}O_2a_{\ve{k}_2}O_3\mid-\ve{k}', \psi_{n'}\ra\ra &&= 
\int d\ve{x}\exp(\imath \ve{k}.\ve{x})\la\la\Psi(n, \ve{x})\mid O_1a^{\dagger}_{\ve{k}}O_2a_{\ve{k}'}O_3\mid-\ve{k}',  \psi_{n'}\ra\ra\eol&&= 
\int d\ve{x}\exp(\imath \ve{k}.\ve{x})\la\la\Psi(n, \ve{x}=0)\mid\exp(\imath \ve{P}.\ve{x}) O_1a^{\dagger}_{\ve{k}_1}O_2a_{\ve{k}_2}O_3\mid-\ve{k}',  \psi_{n'}\ra\ra,\eol&&\label{MEV1}\eeq
where $\ve{P}$ is the momentum operator in Fock space and 
\beq\mid\Psi(n, \ve{x})\ra\ra=\exp(-\imath \ve{P}.\ve{x})\mid\Psi(n, \ve{x}=0)\ra\ra, \label{Ptrans}\eeq
has been used. 

If the $O_i$ are translationally invariant the state appearing to the right of $\ve{P}$ in eq.(\ref{MEV1}) will have momentum $\ve{k}_1-\ve{k}_2-\ve{k}'$. Integrating over $\ve{x}$ eq.(\ref{MEV1}) reduces to
\beq \la\la -\ve{k}, \psi_{n} \mid  O_1a^{\dagger}_{\ve{k}_1}O_2a_{\ve{k}_2}O_3\mid-\ve{k}',  \psi_{n'}\ra\ra&&=(2\pi)^3\delta(\ve{k}+\ve{k}_1-\ve{k}'-\ve{k}_2)\la\la\Psi(n, \ve{x}=0)\mid  O_1a^{\dagger}_{\ve{k}_1}O_2a_{\ve{k}_2}O_3\mid-\ve{k}',  \psi_{n'}\ra\ra.\eol&&\label{MEV2}\eeq
This result gives the matrix element of the special operators (\ref{specOps}) between nonlocalised states in terms of a momentum conserving delta function and a matrix element involving a localised state in the bra. 
 \section{Derivation of the Green function property, eq.(\ref{GEkn}).}\label{Gkn}
This Appendix concerns a property of the set of states $G(E^+)a^{\dagger}_{\ve{k}}\mid- \ve{k}, \psi_n \ra\ra$ with $E$ fixed, $\ve{k},\,n,$ arbitrary and $E^+=E+\imath\epsilon$.

 Using the identity eq.(\ref{Villars1})
    \beq  G(E^+)a^{\dagger}_{\ve{k}}\mid- \ve{k}, \psi_n \ra\ra&&= a^{\dagger}_{\ve{k}}G(E^+-\epsilon_k)\mid- \ve{k}, \psi_n \ra\ra+G(E^+) [V,a^\dagger_{\ve{k}}]_-G(E^+-\epsilon_k)\mid -\ve{k},  \psi_n \ra\ra\eol &&=\frac{1}{(E^+-\frac{(A+1)}{A}\epsilon_k-E_n)}
(a^{\dagger}_{\ve{k}}\mid- \ve{k}, \psi_n \ra\ra+G(E^+)  [V,a^\dagger_{\ve{k}}]_-
\mid -\ve{k},  \psi_n \ra\ra)\eol&&=\frac{1}{(E^+-\frac{(A+1)}{A}\epsilon_k-E_n)}
(a^{\dagger}_{\ve{k}}\mid- \ve{k}, \psi_n \ra\ra+  \frac{1}{2}\int d\ve{k}_3 G(E^+)a^\dagger_{\ve{k}_3}  \mathcal{V}(\ve{k}_3, \ve{k})\mid -\ve{k},  \psi_n \ra\ra),\label{Villars152}\eeq
where eq.(\ref{Vcommidentity223}) has been used to express the commutator $[V,a^\dagger_{\ve{k}}]_-$ in terms of the nucleon number conserving interaction $ \mathcal{V}$ defined in eq.(\ref{Vkk'1}).  Introducing the complete set of $A$ nucleon states $\mid -\ve{k}' , \psi_{n'} \ra\ra$ and using the result (\ref{MEV2}) of Section (\ref{matrixelementsOps}) gives 
\beq G(E^+)a^{\dagger}_{\ve{k}}\mid- \ve{k}, \psi_n \ra\ra&&=\frac{1}{(E^+-\frac{(A+1)}{A}\epsilon_k-E_n)}
(a^{\dagger}_{\ve{k}}\mid- \ve{k}, \psi_n \ra\ra\eol&&+ \int d\ve{k}'\sum_{n'} \, G(E^+)a^\dagger_{\ve{k}'} 
\mid -\ve{k}' , \psi_{n'} \ra\ra\frac{1}{2}\la\la \Psi(n',\ve{x}=0)\mid \mathcal{V}(\ve{k}', \ve{k})\mid-\ve{k},\psi_n\ra\ra)\eol
&&
\label{Villars153}\eeq
Defining  the matrix
\beq \mathcal{V}_{\ve{k}',n';\ve{k},n}=\frac{1}{2} \la\la \Psi(n',\ve{x}=0)\mid\mathcal{V}(\ve{k}', \ve{k})\mid-\ve{k},\psi_n\ra\ra, \label{Vn'k'nk}\eeq
eq.(\ref{Villars153}) can be written
\beq  (E^+-\frac{(A+1)}{A}\epsilon_k-E_n)G(E^+)a^{\dagger}_{\ve{k}}\mid- \ve{k}, \psi_n \ra\ra&&=
a^{\dagger}_{\ve{k}}\mid- \ve{k}, \psi_n \ra\ra\eol&&+\sum_{n'} \int d\ve{k}'\, G(E^+)a^\dagger_{\ve{k}'} 
\mid -\ve{k}' , \psi_{n'} \ra\ra\mathcal{V}_{\ve{k}',n';\ve{k},n}.
\eol &&\label{Villars154}\eeq
This equation can be regarded as an in-homogenous set of couple equations for the ket vectors $G(E^+)a^{\dagger}_{\ve{k}}\mid- \ve{k}, \psi_n \ra\ra$ and can be rewritten as
\beq  \int d{k}'\sum_{n'}G(E^+)a^{\dagger}_{\ve{k}'}\mid- \ve{k}', \psi_{n'} \ra\ra((E^+-\frac{(A+1)}{A}\epsilon_k-E_n)\delta(\ve{k}'-\ve{k})\delta_{n,n'}-\mathcal{V}_{\ve{k}',n';\ve{k},n})&&=
a^{\dagger}_{\ve{k}}\mid- \ve{k}, \psi_n \ra\ra .
\eol &&\label{Villars155}\eeq
A convenient way of expressing the solution of these equations in terms of the inverse of the matrix 
\beq (E^+-\frac{(A+1)}{A}\epsilon_k-E_n)\delta(\ve{k}'-\ve{k})\delta_{n,n'}-\mathcal{V}_{\ve{k}',n';\ve{k},n},\label{EHmatrix}\eeq
 is discussed in Section (\ref{T100}) following eq.(\ref{Tkk'4}).
\section{The identity eq.(\ref{LSZdev22}).}\label{LSZid}
 \beq \la\la\Psi(0,\ve{x}=0)\mid  a^{\dagger}_{\ve{k}'} H&&\!\!\!\!\!\!=\la\la\Psi(0,\ve{x}=0)\mid ( H a^{\dagger}_{\ve{k}'} -[H,a^{\dagger}_{\ve{k}'}])\eol
 &&=\la\la\Psi(0,\ve{x}=0)\mid ( H a^{\dagger}_{\ve{k}'}-\epsilon_{k'} a^{\dagger}_{\ve{k}'}-[V,a^{\dagger}_{\ve{k}'}])\eol
 &&=\la\la\Psi(0,\ve{x}=0)\mid ((\frac{\ve{\ve{P}}^2}{2Am}+E_0)a^{\dagger}_{\ve{k}'}-\epsilon_{k'} a^{\dagger}_{\ve{k}'}-[V,a^{\dagger}_{\ve{k}'}]),\label{LSZdev1}\eeq
where the fact that $\mid\Psi(0,\ve{x}=0)\ra\ra$ is an eigenfunction with  eigenvalue, $E_0$, of the $A$-nucleon intrinsic Hamiltonian, $H-\frac{\ve{P}^2}{2Am}$, has been used in the second line. Within the matrix element in eq.(\ref{LSZdev22}) $\ve{P}$ has the eigenvalue $-\ve{k}$ and eq.(\ref{LSZdev1}) gives 
\beq \la\la\Psi(0,\ve{x}=0)\mid  a^{\dagger}_{\ve{k}'}=\la\la\Psi(0,\ve{x}=0)\mid[V,a^{\dagger}_{\ve{k}'}]\frac{1}{(\frac{\epsilon_{k}}{A}-\epsilon_{k'}+E_0-H)}.\label{LSZdev2}\eeq
 For $A=1$ the bra $\la\la\Psi(0,\ve{x}=0)\mid  a^{\dagger}_{\ve{k}'} \hat {H}$ vanishes and the identity (\ref{LSZdev1}) is not useful. 
 \section{Feshbach theory of the optical potential applied to $B$-space transition operator $\hat{\mathcal{T}}$.}\label{FeshbachTheory}
$\hat{\mathcal{T}}$ is the operator in $B$-space defined for a general $\hat{V}$ by
\beq\hat{\mathcal{T}}(E+\imath \epsilon)&&=\hat{V}+\hat{V} \frac{1}{(E+\imath \epsilon-\hat{T}-\hat{h}_A-\hat{V})}\hat{V}.\eol&&=\hat{V}+\hat{V} \frac{1}{(E+\imath \epsilon-\hat{T}-\hat{h}_A)}\hat{\mathcal{T}}(E+\imath \epsilon).\label{TU}\eeq
The objective of the following algebra is to give uncoupled equations for the matrix elements $P_0\hat{\mathcal{T}}P_0$ and $Q_0\hat{\mathcal{T}}P_0$ where $P_0$ projects on to states $\mid \ve{k}, 0\ra$ in $B$-space in which the target is in its ground state, $n=0$, and $Q_0$ projects on the orthogonal sub-space of states with $n\neq 0.$  For this purpose it is convenient to introduce the operator
\beq \hat{\Omega}(E+\imath \epsilon)=1+\frac{1}{(E+\imath \epsilon-\hat{T}-\hat{h}_A-\hat{V})}\hat{V}, \label{TU2}\eeq
so that eq.(\ref{TU}) can be written
\beq\hat{\mathcal{T}}(E+\imath \epsilon)= \hat{V}\hat{\Omega}(E+\imath \epsilon).\label{TU3}\eeq
Using the properties of $P_0$ and $Q_0$ gives (with $E^+=E+\imath \epsilon$)
\beq(E^+-E_0-\hat{T}-P_0\hat{V}P_0) P_0\hat{\Omega}P_0=(E^+-E_0-\hat{T})P_0+P_0\hat{V}Q_0\hat{\Omega}P_0, \label{TU4}\eeq
and
\beq(E^+-\hat{T}-\hat{h}_A-Q_0\hat{V}Q_0) Q_0\hat{\Omega}P_0=Q_0\hat{V}P_0\hat{\Omega}P_0.\label{TU5}\eeq
Using eq.(\ref{TU5}) to give a formula for $Q_0\hat{\Omega}P_0$ and inserting this into eqs.(\ref{TU4}) and (\ref{TU3})  gives
\beq(E^+-E_0-\hat{T}-P_0\hat{U}_{\mathrm{opt}}P_0) P_0\hat{\Omega}P_0=(E^+-E_0-\hat{T})P_0, \label{TU6}\eeq
and
\beq P_0 \hat{\mathcal{T}}P_0=P_0\hat{U}_{\mathrm{opt}}P_0+P_0\hat{U}_{\mathrm{opt}}P_0\frac{1}{(E^+-E_0-\hat{T}-P_0\hat{U}_{\mathrm{opt}}P_0)} P_0\hat{U}_{\mathrm{opt}}P_0,\label{TU7}\eeq
where 
\beq \hat{U}_{\mathrm{opt}}=\hat{V}+\hat{\mathcal{U}}Q_0\frac{1}{(E+\imath \epsilon-\hat{T}-\hat{h}_A-Q_0\hat{V}Q_0)}Q_0\hat{V}. \label{TU8}\eeq

Note also that according to eq.(\ref{TU6})
\beq P_0\hat{\Omega}P_0=P_0+\frac{1}{(E^+-E_0-\hat{T}-P_0\hat{U}_{\mathrm{opt}}P_0)}P_0\hat{U}_{\mathrm{opt}}P_0.\label{TU9}\eeq
\section{Time-dependent single particle Green's function.}\label{G00t}
When expressed in the time domain the Green's functions relation that is consistent with eq.(\ref{G1n}) is 
\beq
  G_{0,0}(\ve{k},\ve{k}';\omega) &=& \frac{1}{\imath }\int _{-\infty}^{+\infty}\,dt\,\exp( -\imath\omega t)\exp(-\epsilon \mid t \mid))\eol&\times &\exp( \imath\frac{\epsilon_{k'}}{A} t)\la\la\Psi(0,\ve{x}=0)\mid  \mathcal{T}\{ a^\dagger_{\ve{k}'}(t),\,a _{\ve{k}}(t=0)\}\mid  -\ve{k}',  \psi_{0} \ra\ra,\label{onenucleonGreen24}\eeq
where the time-ordering operator $\mathcal{T}$ for fermions is defined by
\beq  \mathcal{T}\{A(t_2),\,B(t_1)\}&=&A(t_2)\times B(t_1),\,\,\mathrm{for}\,\,t_2>t_1,\eol
&=&-B(t_1)\times A(t_2),\,\,\mathrm{for}\,\,t_1>t_2.\label{timeorder2}\eeq
The Heisenberg operators appearing in eq.(\ref{onenucleonGreen24}) are defined by
 \beq  a^\dagger _{\ve{k}}(t)=\exp (\imath H t)a^\dagger _{\ve{k}}\exp (-\imath H t).\label{adaggerHeis2}\eeq

\end{document}